\documentclass[aps,prx,twocolumn,superscriptaddress,longbibliography,floatfix]{revtex4-2}

\usepackage[colorlinks=true,urlcolor=blue,citecolor=blue,allcolors=blue]{hyperref}

\usepackage{color}
\usepackage{braket}
\usepackage{amsmath}
\usepackage{esint}
\usepackage{amssymb}
\usepackage{physics}
\usepackage[utf8]{inputenc}
\usepackage{csquotes}
\usepackage{bm}
\usepackage{graphicx}
\usepackage{epstopdf}
\usepackage{tikz}
\usetikzlibrary{calc}
\usepackage{soul}
\usepackage{xcolor}
\usepackage{bookmark}
\usepackage{array}
\usepackage{slashed}
\usepackage{bbm}
\usepackage{microtype}
\usepackage{upgreek}
\usepackage{ulem}
\usepackage[capitalize]{cleveref}

\renewcommand{\O}{\mathcal{O}}
\newcommand{\eff}{\textrm{eff}}

\renewcommand{\H}{\mathcal{H}}

\newcommand{\V}{\mathcal{V}}

\def\bbra#1{\mathinner{\langle\hspace{-0.75mm}\langle{#1}|}}
\def\kket#1{\mathinner{|{#1}\rangle\hspace{-0.75mm}\rangle}}
\def\updownharpoons{\mathinner{\upharpoonleft\hspace{-0.75mm}\downharpoonright}}
\def\downupharpoons{\mathinner{\downharpoonleft\hspace{-0.75mm}\upharpoonright}}

\newcommand{\corr}[1]{\textcolor{black}{#1}}

\begin{document}
\title{Theory of Multi-photon Processes for Applications in Quantum Control}

\author{Longxiang Huang}
\altaffiliation{These authors contributed equally to this work.}
\affiliation{Technical University of Munich, TUM School of Natural Sciences, Department of Physics, 85748 Garching, Germany}
\affiliation {Walther-Meissner-Institute, Walther-Meissner-Strasse 8, 85748 Garching, Germany}

\author{Jacquelin Luneau}
\altaffiliation{These authors contributed equally to this work.}
\affiliation{Technical University of Munich, TUM School of Natural Sciences, Department of Physics, 85748 Garching, Germany}
\affiliation {Walther-Meissner-Institute, Walther-Meissner-Strasse 8, 85748 Garching, Germany}
\affiliation{Munich Center for Quantum Science and Technology (MCQST), 80799 Munich, Germany}

\author{Johannes Schirk}
\affiliation{Technical University of Munich, TUM School of Natural Sciences, Department of Physics, 85748 Garching, Germany}
\affiliation {Walther-Meissner-Institute, Walther-Meissner-Strasse 8, 85748 Garching, Germany}

\author{Florian Wallner}
\affiliation{Technical University of Munich, TUM School of Natural Sciences, Department of Physics, 85748 Garching, Germany}
\affiliation {Walther-Meissner-Institute, Walther-Meissner-Strasse 8, 85748 Garching, Germany}

\author{Christian M. F. Schneider}
\affiliation{Technical University of Munich, TUM School of Natural Sciences, Department of Physics, 85748 Garching, Germany}
\affiliation {Walther-Meissner-Institute, Walther-Meissner-Strasse 8, 85748 Garching, Germany}

\author{Stefan Filipp}
\affiliation{Technical University of Munich, TUM School of Natural Sciences, Department of Physics, 85748 Garching, Germany}
\affiliation {Walther-Meissner-Institute, Walther-Meissner-Strasse 8, 85748 Garching, Germany}
\affiliation{Munich Center for Quantum Science and Technology (MCQST), 80799 Munich, Germany}

\author{Klaus Liegener}
\affiliation {Walther-Meissner-Institute, Walther-Meissner-Strasse 8, 85748 Garching, Germany}

\author{Peter Rabl}
\affiliation{Technical University of Munich, TUM School of Natural Sciences, Department of Physics, 85748 Garching, Germany}
\affiliation {Walther-Meissner-Institute, Walther-Meissner-Strasse 8, 85748 Garching, Germany}
\affiliation{Munich Center for Quantum Science and Technology (MCQST), 80799 Munich, Germany}

\date{\today}

\begin{abstract}
We present a general theoretical framework for evaluating multi-photon processes in periodically driven quantum systems, which have been identified as a versatile tool for engineering and controlling nontrivial interactions in various quantum technology platforms. To achieve the accuracy required for such applications, the resulting effective coupling rates, as well as any drive-induced frequency shifts, must be determined with very high precision. Here, we employ degenerate Floquet perturbation theory together with a diagrammatic representation of multi-photon processes to develop a systematic and automatable approach for evaluating the effective dynamics of driven quantum systems to arbitrary orders in the drive strength. As a specific example, we demonstrate the effectiveness of this framework by applying it to the study of multi-photon Rabi oscillations in a superconducting fluxonium qubit, finding excellent agreement between our theoretical predictions and exact numerical simulations, 
\corr{even for large driving amplitude.}
\end{abstract}

\maketitle


\section{Introduction}
Multi-photon processes are a hallmark of nonlinear quantum optical systems. While originally studied primarily in the context of multi-photon transitions and ionization effects in atoms and molecules~\cite{abella_optical_1962,bonch-bruevich_multiphoton_1965,bebb_multiphoton_1966,agostini_multiphoton_1970,young_third-harmonic_1971,lambropoulos_topics_1976,gontier_multiphoton_1971,lambropoulos_two-electron_1998,chu2004FloquetTheoremGeneralizeda}, such processes are now widely used as versatile control techniques for quantum technology applications, ranging from multi-photon driving schemes for quantum dots \cite{villafane_three-photon_2023}, spin qubits~\cite{szechenyi_maximal_2014,romhanyi_subharmonic_2015} and superconducting circuits~\cite{oliver_mach-zehnder_2005,shevchenko2012MultiphotonTransitionsJosephsonjunction,nesterov2021ProposalEntanglingGates,xiong2022ArbitraryControlledphaseGate,sah_decay-protected_2024,xia_fast_2025,schirk_subharmonic_2025} to Floquet engineering of many-body interactions~\cite{goldman_periodically_2014, goldman_periodically_2015, bukov_universal_2015,eckardt_high-frequency_2015,eckardt_colloquium_2017, schweizer_floquet_2019,rodriguez-vega_low-frequency_2021}. Compared to resonant control, this approach offers the advantage that novel effective interactions can be mediated via the virtual population of other off-resonant states, and also that the control frequency can be far removed from the relevant system frequencies. This crucial feature breaks the usual trade-off between the ability to apply fast control operations and the requirement to shield the quantum system from its environment. 
\corr{For example, recent studies in superconducting qubits have shown that applying a control signal at one-third of the qubit’s transition frequency enables the shielded drive line to induce decay-protected Rabi oscillations between two states of a transmon~\cite{sah_decay-protected_2024, xia_fast_2025} as well as a heavily shielded fluxonium qubit~\cite{schirk_subharmonic_2025}.}

While in many use cases, the effective dynamics that is generated by a periodic modulation can already be well approximated by second-order perturbation theory, this is clearly no longer possible for the higher-order processes mentioned above. 
More generally,  for quantum control applications, a very precise knowledge of all drive-induced couplings and frequency shifts is required, for which processes beyond the leading order can still give relevant contributions. Therefore, theoretical approaches based on the rotating wave-approximation (RWA)~\cite{sah_decay-protected_2024,xia_fast_2025}, the truncation to a two-level system~\cite{sarkar_subharmonics_2021}, nonlinear resonator Hamiltonians~\cite{xiao2024} or the Floquet-Magnus expansion ~\cite{magnus_exponential_1954,casas2001Floquet,rahav2003Effective,goldman_periodically_2014, goldman_periodically_2015, bukov_universal_2015,eckardt_colloquium_2017, schweizer_floquet_2019,rodriguez-vega_low-frequency_2021,xu_perturbative_2025,schirk_subharmonic_2025} are often not sufficient to reach the required level of accuracy or become impractical to evaluate at higher orders.  

In this paper, we present a general theoretical framework for evaluating multi-photon processes in periodically driven quantum systems, to arbitrarily high levels of precision depending on given numerical computation power. In our approach, we first apply Floquet theory to map the driven quantum system onto a time-independent problem in the extended Sambe space~\cite{shirley_solution_1965,sambe_steady_1973}. In this representation, we then employ degenerate perturbation theory~\cite{soliverez_effective_1969} to obtain the effective system dynamics within a relevant set of quasi-resonant states. Mapping this result back into the physical Hilbert space yields not only the effective coupling rates and frequency shifts to arbitrary order in the drive strength, but also allows us to predict both the slowly evolving and rapidly oscillating components of the system's wave function. This information is essential for accurately predicting the state of the system over time and enabling the design of high-fidelity control methods. A key advantage of this approach is that each order in the perturbation theory can be determined iteratively and in a systematic manner, such that the evaluation of higher-order contributions can be easily automated. Furthermore, each term in the perturbation series can be represented by a diagram for an intuitive physical interpretation.

As two paradigmatic examples, we discuss the application of this perturbation theory for modeling multi-photon Rabi oscillations in a subharmonically driven two-level system and in a fluxonium qubit, motivated by recent experimental demonstrations of such driving schemes with spin qubits and superconducting quantum circuits \cite{romhanyi_subharmonic_2015,xia_fast_2025,schirk_subharmonic_2025}. These examples illustrate that already for modest drive strengths, a precise knowledge of the frequency shifts and contributions to the Rabi frequency beyond leading order is necessary to obtain the accuracy required, for example, for high-fidelity qubit operations. The presented theoretical framework is, however, completely general. To facilitate its adoption for other periodically driven quantum systems, we provide a sample Python code that automates the evaluation of higher-order processes for arbitrary system Hamiltonians and perturbation operators~\cite{ZenodoCode}. The precise knowledge of the effective coupling parameters and frequency shifts obtained in this way can successively be used as input for various optimal control techniques \cite{werninghaus_leakage_2021, glaser_sensitivity-adapted_2024}, which are already routinely applied for resonantly driven systems. As a basic example, we show how the predictions from perturbation theory can be used to design an experimentally more relevant flat-top pulse to implement multi-photon $\pi$-pulses with very high fidelities. This represents a relevant test case, where previous tools such as the RWA fail, mainly due to the significant anharmonicity of the fluxonium and the low drive frequency.

The remainder of the paper is structured as follows. In Sec. \ref{sec:RF-P}, we first lay out the general theoretical framework and introduce the relevant notation and main steps that are involved in the application of degenerate Floquet perturbation theory. The relevance of this framework can already be seen in simple toy models such as the XZ-model, which feature driving due to higher order processes as we outline in Sec. \ref{sec:A_to_SHD}. \corr{Already at this level, a general procedure becomes apparent which can be captured in a diagrammaic framework.} In Sec. \ref{sec:arbitrary_order_expansion} we then discuss in more detail the general results of the degenerate perturbation theory at arbitrary orders and illustrate how it predicts drive rates, Stark shifts and fast oscillations. \corr{After deriving the general effective Hamiltonian on long time-scales, we comment on computing its sub-leading corrections, so-called fast oscillations.} The generality of the equations provided here makes them easily applicable to various scenarios\corr{, which is done in the next two sections}. In Sec. \ref{sec:RabiModel}\corr{, we apply the theory to a two-level system and observe the relevance of higher orders to predict the system accurately even at strong drive powers. Further, in} Sec. \ref{sec:fluxonium}, we \corr{study} the case of a subharmonically driven fluxonium under realistic experimental conditions. \corr{Even in such low-frequency systems with large anharmonicity, the degenerate perturbation theory is able to reach the level of direct numerical simulations, especially when fast oscillations are taken into account.} Finally, in Sec. \ref{sec:conclusion} we present the main conclusions of this work together with an outlook on its use in quantum control applications.

\section{General framework}
\label{sec:RF-P}

\begin{figure}
    \centering
    \includegraphics[width=\columnwidth]{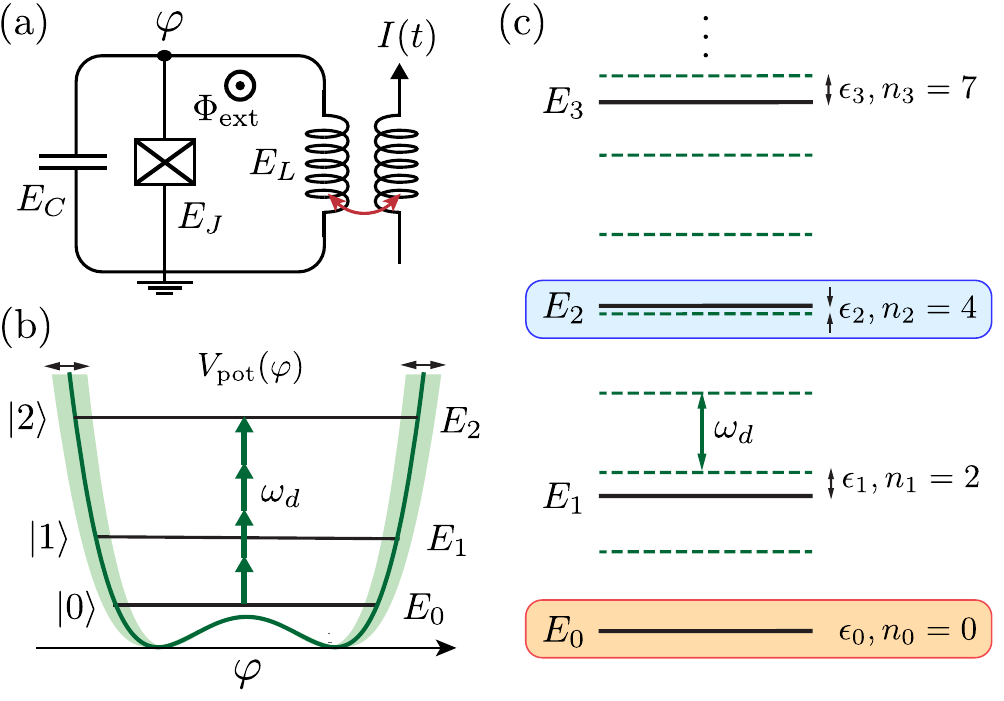}
    \caption{{\bf A multi-photon driven quantum system}.
    (a) Sketch of the circuit of a driven fluxonium qubit, which is used in our analysis as a specific example of a multi-photon driven quantum system. (b) As discussed in more detail in Sec.~\ref{sec:fluxonium} below, the fluxonium qubit is described by a double-well potential $V_{\rm pot} (\varphi)$ for the quantized phase variable $\varphi$, which determines the eigenstates $|k\rangle$ and eigenenergies $E_k$ of this system. Driving the circuit by an external AC current $I(t)\sim \cos(\omega_d t)$ induces additional periodic modulations of the potential, which can be used to implement multi-photon transitions between two or several of the bare eigenstates whenever the frequency difference matches a multiple of the drive frequency $\omega_d$. (c) Identification of a set of quasi-resonant states as discussed in Sec.~\ref{subsec:QuasiResonantSubspace}. 
The black lines represent the energies~$E_k$, which are decomposed according to Eq.~\eqref{eq:energy_decomposition}. The green dashed lines indicate integer multiples of the drive frequency $\omega_d$, added to the energy of the reference state $|0\rangle$. In the example configuration depicted in this plot, the state~$\ket{0}$ is coupled to state $\ket{2}$ via a $n_2=4$ photon process and with a residual detuning of~$|\epsilon_2|\ll\omega_d$. The same level configuration then maps to an exact degeneracy between the eigenstates $\kket{0,0}$ and $\kket{2,4}$ of the extended Hamiltonian $\H_0$ in the Sambe representation used in Eq.~\eqref{eq:FloquetStateOrder0}. }
\label{fig:levels_notations}
\end{figure}

We are interested in the effective dynamics of a periodically driven quantum system, where higher-order processes can induce resonant multi-photon transitions within a given subset of the unperturbed eigenstates. As an illustrative example for such a scenario, Fig.~\ref{fig:levels_notations}(a) shows the circuit of a superconducting fluxonium qubit, which is weakly driven by an external AC current,  $I(t)\sim \cos(\omega_d t)$. As depicted in Fig.~\ref{fig:levels_notations}(b), this drive periodically modulates the potential energy of the quantized phase variable $\varphi$, and induces resonant multi-photon transitions between different eigenstates of the circuit, whenever their energy difference is close to an integer multiple of the driving frequency, $\omega_d$. In order to use such driving schemes for quantum control applications, a precise knowledge of the resulting effective coupling strength as well as the drive-induced frequency shifts is required. 

While we will return to the example of the fluxonium in Sec.~\ref{sec:fluxonium} for a concrete application of our theory,  we first develop the general framework for this analysis for an arbitrary periodically driven quantum system with a Hamiltonian of the form 
\begin{equation}
    H(t) = \sum_k E_k \ket{k}\bra{k} + \bar V(t).
    \label{eq:general_hamiltonian}
\end{equation} 
Here, $\ket{k}$ and $E_k$ denote the eigenstates and the corresponding eigenenergies of the bare system Hamiltonian and $\bar V(t)=\bar V(t+T)$ is a weak periodic perturbation with period $T=2\pi/\omega_d$. $\bar V(t)$ can be expanded in terms of its Fourier components as
\begin{equation}\label{eq:Vbar}
    \bar V(t) = \sum_{p=-\infty}^\infty \bar V_p e^{-ip\omega_d t},
\end{equation}
where $\bar V_p=\bar V_{-p}^\dagger$. With this convention, the components of this expansion with $p>0$ ($p<0$) can be interpreted as the operators associated with the absorption (emission) of $p$ photons from (into) the drive. Note that Eq.~\eqref{eq:Vbar} includes a possible static perturbation with $\bar V_0\neq 0$.

\subsection{Quasi-resonant subspaces} \label{subsec:QuasiResonantSubspace}
Starting from a given reference state $|0\rangle$, for example, the ground state or another state that the system is initialized in,  our goal is to find an effective description of the system's dynamics within a reduced subset of states $|k\rangle \in D$ that are efficiently coupled to $|0\rangle$ via multi-photon processes. To identify these states, we write the bare eigenenergies as 
\begin{equation}\label{eq:energy_decomposition}
    E_k - E_0 = n_k \omega_d + \epsilon_k,
\end{equation}
where $n_k \in \mathbb{Z}$ and $ \epsilon_k \in [-\omega_d/2, \omega_d/2]$. As illustrated by the example shown in Fig.~\ref{fig:levels_notations}(c), in this decomposition $n_k$ is the number of drive photons that matches most closely the energy gap between states $|0\rangle$ and $|k\rangle$ and  $ \epsilon_k$ is the residual frequency detuning from an exact $n_k$-photon resonance. Based on this decomposition, we denote by  
\begin{equation} 
D=\left\{\ket{k} |~ |\epsilon_k| \ll \omega_d\right\}
\end{equation}
the set of states that are quasi-resonant with $|0\rangle$. This set includes the state~$\ket{0}$ with~$\epsilon_0,n_0\equiv 0$ itself.

For a consistent application of the degenerate perturbation theory discussed below, it is necessary to apply the theory  to a set of states with exact multi-photon resonances. To do so, we write the total Hamiltonian as 
\begin{align}
        H(t) &=  H_0 + V(t),
\end{align}
where we have introduced the unperturbed Hamiltonian $H_0 = \sum_{k}\tilde E_k\ket{k}\bra{k}$ with energies 
\begin{align}
    \tilde E_k &= E_k - \epsilon_k = E_0 + n_k\omega_d \quad \text{for}\quad |k\rangle \in D, 
    \label{eq:notation_energy}
\end{align}
and $\tilde E_k = E_k $ otherwise. In turn, the drive term is redefined as 
\begin{align}
     V(t) = \sum_{k\in D}\epsilon_k\ket{k}\bra{k}+ \bar V(t) =  \sum_{p=-\infty}^{\infty} e^{-ip\omega_d t}V_p,
\end{align}
with $ V_p = \bar V_p$ for $p\neq 0$ and 
\begin{equation}
    V_0= \sum_{k\in D}\epsilon_k\ket{k}\bra{k} + \bar V_0.
\end{equation}
Therefore, after this redefinition, $V(t)$ now includes the small residual detunings $\epsilon_k$, which have been omitted from $H_0$, as part of the static perturbation.

\subsection{Sambe representation}\label{sec:Sambe}

In view of the periodicity of the perturbation, it is convenient to use Floquet theory to map $H(t)$ into a time-independent Hamiltonian $\mathcal{H}$ in an extended Hilbert space, which is also known as the Sambe representation~\cite{shirley_solution_1965,sambe_steady_1973}. The mapping can be established by considering \corr{the  Floquet states} $\ket{\psi_s(t)}$ of quasi-energy $\epsilon_s^F$.
\corr{Theses states form a complete basis of solutions of the Schr{\"o}dinger equation on which the initial state can be expanded.}
\corr{The Floquet states are periodic up to a phase, i.e. $\ket{\psi_s(t)}=e^{-i\epsilon_s^F t}\ket{u_s(t)}$, with $\ket{u_s(t+T)}=\ket{u_s(t)}$.
By denoting by $\ket{u_{s,p}}$ the Fourier harmonics of the periodic state~$\ket{u_s(t)}$, the Floquet states are decomposed as
}
\begin{equation}
\ket{\psi_s(t)}=e^{-i\epsilon_s^F t}\sum_{p=-\infty}^\infty e^{-ip\omega_d t}\corr{\ket{u_{s, p}}},
\end{equation} 
\corr{where} the harmonics \corr{$\ket{u_{s, p}}$} can be formally 
grouped into an extended  column vector
\begin{equation}
\label{eq:Floquet_state_vector}
    \kket{\Phi_s} \equiv 
    \begin{pmatrix}
        \vdots \\
        \corr{\ket{u_{s,-1}}} \\
        \corr{\ket{u_{s,0}}} \\
        \corr{\ket{u_{s,+1}}} \\
        \vdots
    \end{pmatrix}.
\end{equation}
This vector is then an eigenstate of the extended Floquet-Sambe Hamiltonian, i.e., 
\begin{equation}
  \mathcal{H}  \kket{\Phi_s}=(\H_0 + \mathcal{V})\kket{\Phi_s} =\epsilon_s^F\kket{\Phi_s}.
\end{equation}
Here, the bare Hamiltonian, 
\begin{align}
    \H_0 & =
    \begin{pmatrix}
        \ddots & & & & & & \\
        & H_0 +2 \omega_d & & & & \\
        & & H_0 +\omega_d & & & &\\
        & & & H_0 & & & \\
        & & & &  H_0 - \omega_d & & \\
        & & & & & & \ddots 
    \end{pmatrix},
\end{align}
is block-diagonal with each block being shifted in energy by a multiple of $\omega_d$. The eigenstates of $\H_0$ in this extended space are
\begin{equation}
    \kket{k,p}=\begin{pmatrix}
        \vdots \\
        0 \\
        \ket{k} \\
        0 \\
        \vdots
    \end{pmatrix}
    \leftarrow p\text{-th row},
    \label{eq:FloquetStateOrder0}
\end{equation}
and have an energy of $\tilde E_k- p\omega_d$. Therefore, according to the convention introduced in Eq.~\eqref{eq:notation_energy}, the states $\kket{k, p=n_k}$ are degenerate for~$\ket{k}\in D$.

The periodic perturbation couples different harmonics of the state decomposition in the original frame and therefore maps onto an off-diagonal perturbation of the form
\begin{align}
    \mathcal{V} 
    &=
    \begin{pmatrix}
         \ddots  &  \ddots &  \ddots & &\\
          \ddots& V_0 & V_{-1} & V_{-2} &\\
        \ddots& V_{1} & V_0 & V_{-1} &\ddots \\
        & V_{2} & V_{1} &  V_0 & \ddots \\
        & &  \ddots &  \ddots &\ddots 
    \end{pmatrix} \label{eq:VSambe}
\end{align}
in the extended Sambe space. The relevant matrix elements,
\begin{align}\label{eq:notation_V_elements}
V_{p_1-p_2, k_1 k_2}\equiv 
\bbra{k_1,p_1}\V\kket{k_2,p_2}
= 
\bra{k_1}V_{p_1-p_2}\ket{k_2},
\end{align}
reduce to the matrix elements of the respective harmonics of the original perturbation operator. 

Note that the Sambe representation is closely related to a quantum description of the drive~\cite{shirley_solution_1965}.
In that sense, it is useful to think about~$\kket{k,p}$ as the system being in state~$\ket{k}$ and the drive in the Fock state of~$n_{\text{ref}}-p$ photons, with~$n_{\text{ref}}$ a large but otherwise arbitrary reference photon number. In such a picture, the degeneracies appear more naturally by taking the energy $\omega_d$ of each of the absorbed or emitted photons into account. 
The Sambe representation also shares similarities with the dressed-atom representation used to study multi-photon emission of a driven two-level system~\cite{rudolph_multiphoton_1998,rudolph_multiphoton_1998-1,rudolph_shift_1998,munoz2014EmittersNphotonBundles,munoz2018FilteringMultiphotonEmission}.

\begin{figure*}
	\includegraphics[width=2\columnwidth]{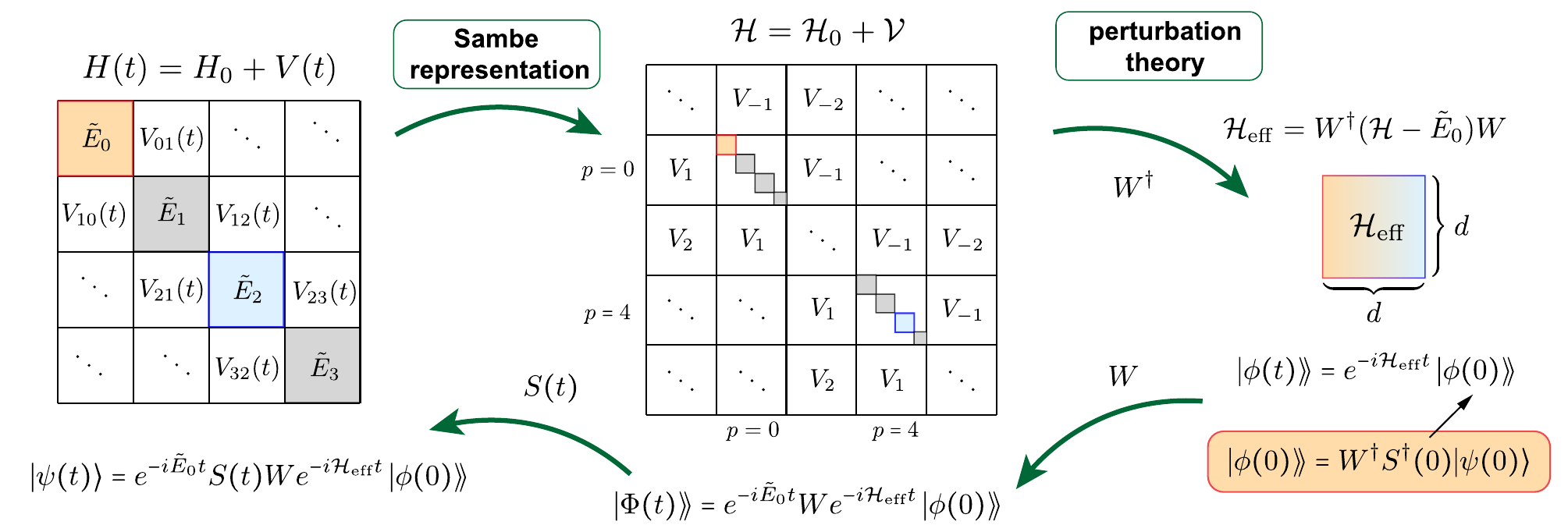}
	\caption{{\bf Degenerate Floquet perturbation theory.} This figure summarizes the general theoretical framework used for the description of a multi-photon-driven quantum system. The starting point is the time-dependent Hamiltonian, for which we identify the relevant set of quasi-resonant states $|k\rangle \in D$ (indicated in orange and blue) with energies $\tilde E_k=\tilde E_0+n_k \omega_d$. This Hamiltonian is then mapped onto a time-independent Hamiltonian $\H$ in the extended Sambe space, where the states $|k\rangle \in D$ are mapped onto degenerate unperturbed states $\kket{k,n_k}$ located in different photon-number sectors (in orange \corr{and blue}).
    Finally, degenerate perturbation theory is used to find a transformation $W^\dag$ such that the dynamics of this degenerate subspace is decoupled from the other states and described by an effective Hamiltonian $\H_\eff$. Once the evolution in this transformed subspace has been obtained, it can be mapped back onto the corresponding time-evolved states in the full Sambe space and, successively, onto the time-evolved state $|\psi(t)\rangle$ in the physical Hilbert space. }
\label{Fig:2}
\end{figure*}

\subsection{Degenerate Perturbation Theory}\label{sec:DPT}
As already mentioned above, in the absence of the perturbation, 
the states $\kket{k,n_k}$ with $\ket{k}\in D$  obey 
\begin{align}
    \H_0\kket{k, n_k} &= \tilde E_0 \kket{k,n_k},
\end{align}
and form a $d$-dimensional degenerate subspace with eigenvalue $\tilde E_0$. In the following, we denote by 
\begin{align}\label{eq:P_projector}
    P = \sum_{k\in D}\kket{k,n_k}\bbra{k,n_k},
\end{align}
the projector onto this subspace. All other states $\kket{l,p}$ with either $\ket{l}\notin D$ or $p\neq n_l$ span the complementary subspace with projection operator $Q=1-P$. States in this subspace obey  
\begin{align}
     \H_0\kket{l,p} &= (\tilde E_l-p\omega_d) \kket{l,p}
\end{align}
and are thus separated from the degenerate subspace of interest by a finite energy gap. 

The perturbation $\V$ couples the two sets of bare eigenstates, but as long as the drive strength is sufficiently weak, we can use degenerate perturbation theory to evaluate its effect on the states in the $P$-subspace in a systematic manner. In the current setting, such a perturbative treatment is applicable as long as the condition  
\begin{equation}
\label{eq:static_resonance_condition}
|\bar V|, |\epsilon_k| \ll \omega_d, |\epsilon_l|,
\end{equation}
where $|\bar V|=\max_{p,k,l}|\bar V_{p,kl}|$, is satisfied for all relevant states $\ket{k}\in D$ and $\ket{l}\notin D$. Formally, we can introduce a dimensionless perturbation parameter $\lambda$ by writing $|\bar V|/\omega_d,|\epsilon_k|/\omega_d=\mathcal{O}(\lambda)$. Then, a perturbative treatment of $\V$ is valid when $\lambda\ll 1$, and $\mathcal{O}(\lambda^r)$ denotes a term that is of $r$-th order in this small parameter. 
Note that different types of multi-photon transitions can occur when the driving strength is larger than the driving frequency. For a two-level system, those processes can be captured by the Landau-Zener-St\"uckelberg approach~\cite{shevchenko2010LandauZenerStuckelbergInterferometry,ivakhnenko2023NonadiabaticLandauZener}.


As illustrated in Fig.~\ref{Fig:2}, degenerate perturbation theory allows us to derive \corr{a transformation $W$ to} an effective Hamiltonian 
\begin{equation}\label{eq:Heff_definition}
\H_\eff = W^\dag (\H-\tilde E_0) W,
\end{equation} 
which acts on the degenerate subspace only, i.e. $P\H_\eff P=\H_\eff$, but whose eigenvalues are---up to a constant offset---those of the perturbed Hamiltonian~$\H$. This means that for the set of exact eigenstates $\kket{\Phi_s}$ of the full Sambe Hamiltonian, 
\begin{equation}
    \H\kket{\Phi_s} =  (\tilde E_0+\delta E_s)\kket{\Phi_s},
\end{equation}
with $s=1,\dots,d$ and $\delta E_s(\lambda\rightarrow 0)=0$, 
the transformed states $\kket{\phi_s}= W^\dag \kket{\Phi_s}$ satisfies 
\begin{equation}
   \H_\eff \kket{\phi_s} = \delta E_s \kket{\phi_s} 
\end{equation}
and lie fully within the degenerate subspace, i.e., $Q\kket{\phi_s}= 0$. Note that $W^\dag$ is not a unitary transformation, as it maps states from the full Sambe space into the reduced $d$-dimensional degenerate subspace, $W^\dag W = P$.

Under the assumption that $\lambda\ll 1$, approximate expressions for $\H_\eff$ and $W$ can be derived  by expanding both operators in powers of $\V$~\cite{soliverez_effective_1969}. The main result of this calculation, which is discussed in more detail in the following sections, is an expansion of  the effective Hamiltonian of the form
\begin{equation}
    \H_\eff = \lim_{r_{H}\to\infty}\H^{[r_{H}]}_\eff,\quad\H^{[r_{H}]}_\eff=\sum_{r=1}^{r_{H}} \H_\eff^{(r)},
\end{equation}
where $\H_\eff^{(r)}=\O(\lambda^r)$ and $r_{H}$ refers to the maximal order to which the perturbation theory is carried out. 
\corr{In this way, the effective Hamiltonian~$\H_\eff$ and the transformation~$W$ can be expressed systematically in terms of the projection~$P$, the perturbation~$\V$, and a so-called resolvent operator, see Appendix~\ref{app:DPT} for the detailed expressions.}
Within the degenerate subspace,  $\H_\eff$ then describes the effective couplings  
\begin{equation}
    \Omega_{lk}=\bbra{l,n_l}\H_\eff\kket{k,n_k}
\end{equation}
between any pair of states as well as the effective energies (or Stark shifts), 
\begin{equation}
    \delta_{k}\equiv\Omega_{kk}=\bbra{k,n_k}\H_\eff\kket{k,n_k},
\end{equation}
of the states. In the following we denote by  $\delta_k^{(r)}=\bbra{k,n_k}\H_\eff^{(r)}\kket{k,n_k}$ and $\Omega_{lk}^{(r)}=\bbra{l,n_l}\H_\eff^{(r)}\kket{k,n_k}$ the same quantities computed at order~$r$.


In many cases of interest, the drive addresses a specific transition, such that the relevant set of resonant states reduces to only two states, which we can label by~$\ket{0}$ and $\ket{1}$. The effective Hamiltonian,
\begin{equation}\label{eq:Heff_2x2}
    \H_{\eff} = \begin{pmatrix}
        \delta_0 & \Omega_{10}^* \\
        \Omega_{10} & \delta_1
    \end{pmatrix},
\end{equation}
then reduces to the well-known Hamiltonian of a driven two-level system with a detuning $\Delta=\delta_1-\delta_0$ and a Rabi frequency $\Omega_R=2|\Omega_{10}|$ on resonance.

\subsection{State evolution}\label{sec:eff_dynamics}
While the effective Hamiltonian~$\H_\eff$, when evaluated to sufficiently high order, provides an accurate approximation of the relevant system energies, it describes the dynamics of a transformed subset of states in the Sambe representation. In most cases, however, we are interested in the evolution of the state vector in the original basis. More specifically, given the initial state 
\begin{equation}\label{eq:initial_state}
    \ket{\psi(t=0)}=\sum_{k\in D}c_k(0)\ket{k}
\end{equation}
before the drive is switched on, we are interested in the evolution of the state
\begin{equation}\label{eq:State_OriginalBasis}
    \ket{\psi(t)}=\sum_{l}c_l(t)\ket{l}
\end{equation}
for times $t>0$. Note that in general this evolution will also contain nonvanishing amplitudes $c_l(t)\neq 0$ for states with $l\notin D$.

As illustrated in  Fig.~\ref{Fig:2}, for a given initial state $\kket{\phi (0)}$ within the $P$-subspace, we can use $\H_\eff$ to evaluate its effective dynamics and apply the transformation $W$ to obtain the corresponding evolution in the full Sambe space, 
\begin{equation}\label{eq:P_to_Sambe} 
\kket{\Phi(t)} =  e^{-i\tilde E_0 t} W e^{-i \H_\eff t} \kket{\phi (0)}. 
\end{equation}
We further introduce the operator
\begin{equation}
S(t)= \sum_{k,p}    e^{-i p \omega_d t}  |k\rangle \bbra{k,p},
\end{equation}
which maps a given quantum state $\kket{\Phi(t)}$ in the Sambe representation, back onto the corresponding time-dependent state $|\psi(t)\rangle$ in the original Hilbert space, i.e., 
\begin{equation}\label{eq:kket_to_ket} 
\ket{\psi(t)}= S(t) \kket{\Phi(t)}.
\end{equation}
By evaluating Eq.~\eqref{eq:P_to_Sambe}  and Eq.~\eqref{eq:kket_to_ket} at time $t=0$, we also obtain the mapping $|\psi(0)\rangle= S(0) W \kket{\phi(0)} $ between the initial states. After inverting this relation~\footnote{
   Note that the eigenstates $\kket{\phi_s}$ of the effective Hamiltonian map to Floquet states $\ket{\psi_s(t)}=e^{-i\corr{\epsilon_s^F} t}S(t)W\kket{\phi_s}$ of quasi-energy $\corr{\epsilon_s^F}$. By Floquet's theorem, these Floquet states are orthonormal if their quasi-energies are not equal up to a multiple of $\omega_d$, which is here guaranteed by the condition of validity of the perturbation theory.
   The orthonormality at $t=0$ gives $\bbra{\phi_s}W^\dagger S^\dagger(0)S(0)W\kket{\phi_{s'}}=\delta_{s,s'}$, i.e. $W^\dagger S^\dagger(0)S(0)W=P$, from which we can invert the initial state relation, $\kket{\phi(0)}=W^\dagger S^\dagger(0)\ket{\psi(0)}$.
}, we end up with the following effective evolution of the system's wavefunction 
\begin{equation}\label{eq:evol_state_allOrders}
\ket{\psi(t)} =     e^{-i\tilde E_0 t}    S(t) We^{-i \H_\eff t}W^\dagger S^\dag (0)  \ket{\psi(0)},
\end{equation} 
or, equivalently, with the following explicit expression for the amplitudes,
\begin{align}
    c_l(t)
    &= \sum_{k,p,p'} e^{-i(E_0+ p \omega_d) t} \bbra{l,p} We^{-i \H_\eff t}W^\dagger \kket{k,p'} c_k(0).\label{eq:evol_amplitude_allOrders}
\end{align}
We see that apart from the effective Hamiltonian, the evaluation of the resulting system dynamics also requires knowledge of the transformation $W$, which can again be obtained from a series expansion of the form 
\begin{equation}
   W = \lim_{r_{W}\to\infty} \corr{W^{[r_{W}]}},\quad \corr{W^{[r_{W}]}}=\sum_{r=0}^{r_{W}} W^{(r)},
\end{equation}
where $W^{(r)}=\O(\lambda^r)$. Note that in general the maximal order $r_{W}$ in the expansion of $W$ can be chosen differently from $r_H$. 

From Eq.~\eqref{eq:evol_amplitude_allOrders} we see that nonvanishing matrix elements between states $\kket{k,n_k}$ and $\kket{l,p}$ with $l\notin D$ lead to transitions out of the degenerate subspace of interest, which we refer to as leakage. A nonvanishing coupling to a state $\kket{k',p\neq n_{k'}}$ with $k^\prime \in D$ does not lead to leakage, but results in rapidly oscillating contributions to the otherwise slow evolution. We emphasize that those contributions are not simply an effect of imperfect initial state preparation or of an abrupt switching on of the drive discussed above. They are a consequence of the fast oscillations of the Floquet eigenstates, which have components on multiple $p$ harmonics, independently of the initial state.

Note that due \corr{to the} restricted number of states that are included in the set of quasi-resonant states, $D$, the results in Eq.~\eqref{eq:evol_state_allOrders} and Eq.~\eqref{eq:evol_amplitude_allOrders} are restricted to initial states that satisfy $P_D\ket{\psi(0)}=\ket{\psi(0)}$, where $P_D= S(0)WPW^\dag S^\dag (0)$ is the projector on the subspace that is covered by our perturbation theory.
The dynamics of the component $(1-P_D)\ket{\psi(0)}$ of the initial state on the complementary subspace can in principle be obtained by computing similarly the dynamics originating from the effective Hamiltonian of the other states $\ket{l}\notin D$. Alternatively, some of the leakage states can be included in $D$ to get an improved effective Hamiltonian in a slightly larger Hilbert space.


\section{Interlude: Two-photon $XZ$ model}
\label{sec:A_to_SHD}
Before we proceed with our general analysis, we first present in this section the two-photon $XZ$ model as a simple example, where many features of multi-photon transitions can already be understood at the level of first- and second-order perturbation theory.
This model is found, for example, in electronically or acoustically driven spin qubits~\cite{romhanyi_subharmonic_2015,cornell_2025,cornell_all-mechanical_2025}, as well as in the parametrically driven fluxonium off sweetspot or the symmetry-broken two-photon process in superconducting circuits~\cite{deppe_two-photon_2008}.

The $XZ$ model describes a two-level system that is driven both in a transverse and longitudinal direction,
\begin{align}\label{eq:Ham_XZ}
    H_{XZ}(t)=\frac{\omega_{01}}{2}\sigma_z + 2\Omega_z\cos(\omega_d t)\sigma_z + 2\Omega_x \cos(\omega_d t)\sigma_x,
\end{align}
where $\omega_{01}$ is the bare transition frequency and $\Omega_{x,z}$ the driving strengths. It reduces to the usual resonantly driven Rabi model when $\Omega_z\approx 0$ and $\omega_d\approx \omega_{01}$. Here we are interested in the two-photon resonance $\omega_d\simeq \omega_{01}/2$, such that  $n_0=0$, $n_1=2$, and $D=\{\ket{0},\ket{1}\}$, according to our conventions.  The only non-zero harmonics of the drive operator are $\bar V_1=\bar V_{-1}=\Omega_z\sigma_z + \Omega_x\sigma_x$, and  $\epsilon_0=0$ and $\epsilon_1=\omega_{01}-2\omega_d$ are the two detunings.

\subsection{First- and second-order Hamiltonian}
The combined effect of the longitudinal and transverse drive in $H_{XZ}(t)$ induces quasi-resonant two-photon transitions between the states $|0\rangle$ and $|1\rangle$. These can be understood at the level of conventional first- and second-order perturbation theory, applied in the Sambe representation for the states $\kket{0,0}$ and $\kket{1,2}$. From the first order effective Hamiltonian, $\H_\eff^{(1)}=P\V P$, 
we obtain 
\begin{align}\label{eq:eff_energy_order1}
    \delta_k^{(1)} &= V_{0,kk} = \epsilon_k,
\end{align} 
since in this model there  is no static \corr{perturbation}, $\bar V_{0,kk}=0$. We also see that there is no first-order effective coupling, 
\begin{align}
    \Omega_{10}^{(1)} &= V_{n_1-n_0,10}=0,
\end{align}
because no driving is applied at the second harmonics, $V_{2,10}=0$.  

At second order in the perturbation, we obtain the effective Hamiltonian $\H_\eff^{(2)}=P\V R\V P$, where $R=Q\frac{1}{E_0-\H_0}Q$ is the resolvent operator (see the general discussion in Sec.~\ref{sec:arbitrary_order_Heff} and Appendix~\ref{app:DPT} below).  
For the two-photon XZ model, this expression can be readily evaluated to obtain the effective coupling rate 
\begin{align}\label{eq:Rabi_order2_XZ}
    \Omega_{10}^{(2)} &= 
    \frac{V_{1,11}V_{1,10}}{\tilde E_0+\omega_d-\tilde E_1} +  \frac{V_{1,10}V_{1,00}}{\tilde E_0+\omega_d-\tilde E_0} \nonumber \\
    &= \frac{\Omega_z\Omega_x}{-\omega_d}+\frac{\Omega_x(-\Omega_z)}{\omega_d}=-2\frac{\Omega_x\Omega_z}{\omega_d}.
\end{align}
We see that there are two different processes that couple the states $\kket{0,0}$ and $\kket{1,2}$, which interfere constructively. Similarly, we obtain the second-order Stark shift
\begin{align}\label{eq:Stark_order2_XZ}
    \delta_{0}^{(2)} = &  \,   \frac{V_{-1,01}V_{1,10}}{\tilde E_0+\omega_d-\tilde E_1} +  \frac{V_{1,01}V_{-1,10}}{\tilde E_0-\omega_d-\tilde E_1}   \nonumber \\ 
   & +  \frac{V_{-1,00}V_{1,00}}{\tilde E_0+\omega_d-\tilde E_0} +  \frac{V_{1,00}V_{-1,00}}{\tilde E_0-\omega_d-\tilde E_0} \nonumber \\ 
    =& -\frac{\Omega_x^2}{\omega_d}-\frac{\Omega_x^2}{3\omega_d}+\frac{\Omega_z^2}{\omega_d}-\frac{\Omega_z^2}{\omega_d}
    =-\frac{4\Omega_x^2}{3\omega_d}
\end{align}
and $\delta_{1}^{(2)}=-\delta_{0}^{(2)}$. Here we obtain already four possible contributions, out of which the two terms $\sim \Omega_z^2$ cancel exactly. Note that the result in Eq.~\eqref{eq:Stark_order2_XZ} goes beyond the usual RWA and accounts for  processes $\sim V_{-1,10}$, which are associated with the simultaneous excitation of the two-level system and the emission of a photon.

\begin{figure}
    \centering
    \includegraphics[width=\linewidth]{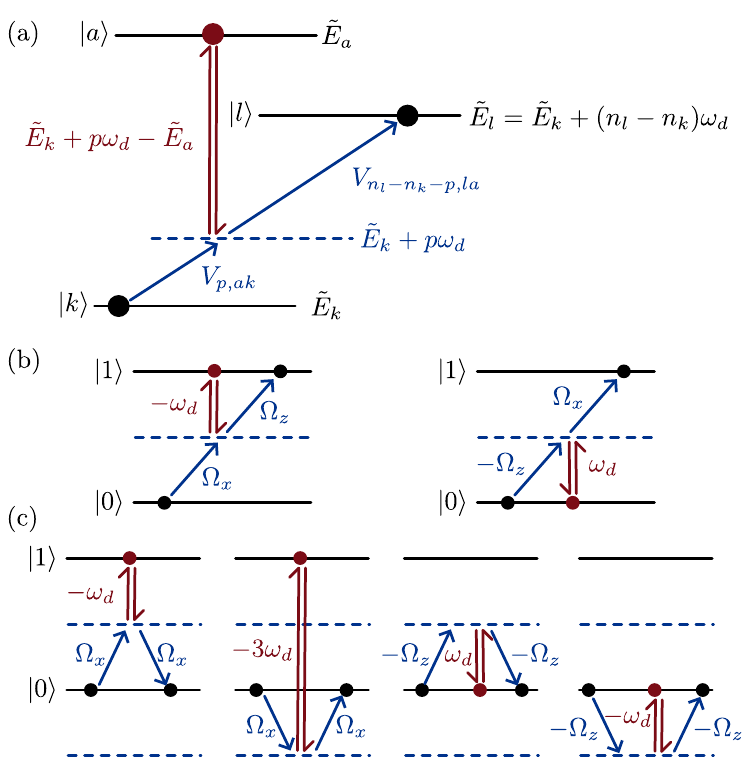}
    \caption{{\bf Diagrammatic representation.} (a) The general diagram representing a second-order process with initial and final states $\ket{k}$ and $\ket{l}$ (black dots), and involving the absorption of~$n_l-n_k$ photons in two steps. 
    The first step is an absorption of~$p$ photons with a virtual transition to the state~$\ket{a}$ (red dot), which is proportional to the matrix element~$V_{p,ak}$.
    The second step is the transition from the virtual state~$\ket{a}$ to the final state~$\ket{l}$ with an absorption of the remaining~$n_l-n_k-p$ photons and with matrix element~$V_{n_l-n_k-p,la}$.
    The whole process is suppressed by an energy denominator (red arrow) that is the difference $\tilde E_k + p\omega_d-\tilde E_a$ between the virtual state energy and the energy of the system after the first step (blue dashed line).
    (b) Diagrams representing the second-order couplings given in Eq.~\eqref{eq:Rabi_order2_XZ} for the two-photon $XZ$-model. Both contributions describe the successive absorption of two individual photons, $p=1$, but with either $\ket{a}=\ket{0}$ or $\ket{a}=\ket{1}$ as the virtual intermediate state.
    (c) Diagrams representing the second-order Stark shift given in Eq.~\eqref{eq:Stark_order2_XZ} for the two-photon $XZ$-model.
    In total four processes contribute to the shift $\delta_0^{(2)}$, in which photons are virtually absorbed ($p=1$) or emitted ($p=-1$). Since the longitudinal coupling does not change the initial state, the two processes with $|a\rangle=|0\rangle$ have energy denominators with exactly opposite sign and cancel each other.
    }
    \label{fig:diagr_order2_general}
\end{figure}

\subsection{Diagrammatic representation}
To generalize the results in Eq.~\eqref{eq:Rabi_order2_XZ}  and Eq.~\eqref{eq:Stark_order2_XZ} to higher orders and higher dimensional systems, it is helpful to represent the individual terms in this expansion graphically.  To do so, we consider the generic second-order contribution 
\begin{align}\label{eq:Rabi_order2_general}
\Omega_{lk}^{(2)} = \dots  + \frac{V_{n_l-n_k-p,la}V_{p,ak}}{\tilde E_k+p\omega_d-\tilde E_a} + \dots,
\end{align} 
which appears in the expansion of the effective coupling between states $\kket{k,n_k}$ and $\kket{l,n_l}$. As illustrated in the corresponding diagram in Fig.~\ref{fig:diagr_order2_general}(a), it can be interpreted as the following two-step process.
In the first step, the system absorbs (or emits)~$p$ photons, leading to an energy increase represented by the first blue arrow. The absorption is accompanied by a virtual transition to an intermediate state~$\ket{a}$ with energy $\tilde E_a$ (red up-arrow), that is coupled to $\ket{k}$ by a nonvanishing matrix element $V_{p,ak}$ of the $p$-th harmonics.
The second step, associated with the matrix element~$V_{n_l-n_k-p,la}$,  describes the absorption (or emission) of the remaining $n_l-n_k-p$ photons accompanied by the transition from $\ket{a}$ to the final state~$\ket{l}$. In the diagram in Fig.~\ref{fig:diagr_order2_general}(a),  this step is represented by the red down-arrow combined with the second blue arrow. 
According to second-order perturbation theory, this two-step transition is suppressed by the difference between the energy $\tilde E_a$ of the virtual state and the energy ~$\tilde E_k+p\omega_d$ of state $|k\rangle$ with $p$ photons added (or subtracted).
The length and direction of the red double arrow represent the absolute value and sign of this energy denominator, respectively, where~$\downupharpoons$ corresponds to a positive sign (the energy of the initial state is higher than the virtual state energy) and~$\updownharpoons$ to a negative sign (the energy of the initial state is lower than the virtual state energy). The whole process can then be visualized by a closed line that follows both the blue and the red arrows. 

In Fig.~\ref{fig:diagr_order2_general}(b) and (c), we apply this diagrammatic representation for the case of the $XZ$ model. We see that all the terms for the second-order effective coupling in Eq.~\eqref{eq:Rabi_order2_XZ}  and the Stark shift in Eq.~\eqref{eq:Stark_order2_XZ}, including their magnitude and sign, can be readily read off from these diagrams, which also provide an intuitive physical interpretation. In Sec.~\ref{sec:RabiHigherOrder}, we provide further examples that show how these diagrammatic prescriptions can be used to visualize also more involved expressions that occur at higher orders of the perturbation.

\subsection{State evolution}\label{sec:evol_XZ}
The example of the two-photon $XZ$ model also allows us to highlight two important issues that are particularly relevant for a precise control of multi-photon driven quantum systems. The first point concerns the identification of the resonance condition $\delta_1=\delta_0$ in the effective Hamiltonian $\H_\eff$, which is required to drive ideal multi-photon Rabi-oscillations between the states $|0\rangle$ and $|1\rangle$. To achieve this condition, we must find the resonant driving frequency $\omega_d^{\text{res}}$, which is the solution of 
\begin{equation}\label{eq:ResonanceCondition} 
\delta_1(\omega_d^{\text{res}})= \delta_0(\omega_d^{\text{res}}). 
\end{equation}
For the $XZ$ model and up to second-order in the perturbation, this condition can still be solved analytically and we obtain
\begin{align}
    \omega_d^{\text{res}}=\frac{\omega_{01}}{4}+\sqrt{\left(\frac{\omega_{01}}{4}\right)^2+\frac{4}{3}\Omega_x^2}.
\end{align}
At higher orders, the resonance conditions can be solved numerically, using, for example, root-finding algorithms, for which the analytic second-order shift can be used as an informed initial guess.

Under the exact resonance condition in Eq.~\eqref{eq:ResonanceCondition}, the time-evolution operator associated with the effective Hamiltonian~$e^{-i\H_\eff t}$ induces complete Rabi oscillations between the states $\kket{0,n_0}$ and $\kket{1,n_1}$. However, as discussed in Sec.~\ref{sec:eff_dynamics} and illustrated in Fig.~\ref{Fig:2}, these Rabi oscillations do not yet represent the evolution of interest, which is the state of the system,  $\ket{\psi(t)}=c_0(t)\ket{0}+c_1(t)\ket{1}$, in the original frame. Therefore, our second task is to evaluate the transformation operator $W$, which connects the two representations. For example, by expanding $W$ in Eq.~\eqref{eq:evol_amplitude_allOrders} up to first order and for an initial state $\ket{\psi(t=0)}=\ket{0}$, we obtain the following evolution [see Sec.~\ref{sec:eff_dynamics_pertb}] 
\begin{align}\label{eq:cl_XY}
    c_1(t) \simeq &\,  e^{-i(\corr{-}\frac{\omega_{01}}{2}+2\omega_d\corr{+\delta_0^{(2)}})t}
    \bigg[\nonumber\\
    &-\left(i + 2\frac{\Omega_z}{\omega_d}\sin(\omega_d t) \right) \sin\left(\Omega_{10}^{(2)}t\right) \nonumber\\
    &+\frac{\Omega_x}{\omega_d}\left(\frac{4}{3}-e^{i\omega_d t}-\frac{e^{3i\omega_d t}}{3}\right)\cos\left(\Omega_{10}^{(2)}t\right)
    \bigg]
\end{align}
for the amplitude of state $|1\rangle$. We see that, on top of the usual Rabi-oscillations, which implement a complete $\pi$-rotation at a time $t_\pi=\pi/(2|\Omega_{10}^{(2)}|)$, there are additional fast oscillations that become increasingly relevant as the drive strength is increased. In general, these fast oscillations prevent a complete state flip at the expected time $t_\pi$, and must be carefully taken into account for precise control operations.

\section{Arbitrary-order expansion}\label{sec:arbitrary_order_expansion}
Returning to the general case, in this section we detail how the effective Hamiltonian and effective state evolution can be systematically evaluated at arbitrary order in perturbation theory. Beyond second order, this task becomes increasingly involved, as the number of transitions to all possible intermediate states increases exponentially and additional processes can occur that have a vanishing energy denominator and must be treated separately. Nevertheless, as shown in Appendix~\ref{app:DPT}, for a given order $r$, closed expressions for $\H_\eff$ and the transformation $W$ can be obtained in a systematic manner. This analysis follows the general framework for degenerate perturbation theory discussed in Ref.~\cite{soliverez_effective_1969}, but is specifically adapted to the Sambe representation used in this work.

\subsection{Effective Hamiltonian}\label{sec:arbitrary_order_Heff}
We first provide the general expression of the effective Hamiltonian and the simplified expression of its leading order in the most common case of a monochromatic drive.
%
%
As shown in Appendix~\ref{app:DPT}, for a given order $r$, the closed expressions for the coupling rates $\Omega_{lk}^{(r)}$ and the Stark shifts $\delta_k^{(r)}\equiv\Omega_{kk}^{(r)}$ are given by
\begin{widetext}
    \begin{align}\label{eq:Heff_matEl_general}
        \Omega_{lk}^{(r)}
        =
        \sum_{\substack{ p_1,\dots,p_r \\ m_1,\dots,m_{r-1} \\ |a_1\rangle, \dots,|a_{r-1}\rangle}} 
        \frac{ c_{m_{r-1},\dots,m_{1}}V_{p_r,la_{r-1}}V_{p_{r-1},a_{r-1}a_{r-2}}\dots V_{p_2,a_2 a_1}V_{p_1,a_1 k}}{(\tilde E_k+(p_{r-1}+\dots+p_1)\omega_d-\tilde E_{a_{r-1}})^{m_{r-1}}\dots(\tilde E_k+(p_2+p_1)\omega_d-\tilde E_{a_2})^{m_2}(\tilde E_k+p_1\omega_d-\tilde E_{a_1})^{m_1}},
    \end{align}
\end{widetext}
where the sum satisfies the following rules:
\begin{enumerate}
    \item The harmonics $p_1,\dots,p_r\in\mathbb{Z}$ correspond to the number of photons absorbed at each step. They are constrained by the condition
    \begin{equation}\label{eq:harmonicsSum}
        p_1+\dots+p_r = n_l-n_k
    \end{equation}
    to ensure the correct total number of photons is absorbed.
    \item The $r-1$ virtual states $\ket{a_1},\dots,\ket{a_{r-1}}$ are any states of the system.
    \item The exponents $m_1,\dots,m_{r-1}\in \mathbb{N}$ satisfy 
    \begin{equation}
        m_1 + \dots + m_{r-1} = r-1
    \end{equation}
    and are zero for resonant intermediate states (if any) and non-zero otherwise:
    \begin{align}
        \tilde E_{a_j}&=\tilde E_k+(p_j+\dots+p_1)\omega_d \quad \Rightarrow \quad m_j=0, \label{eq:constraint_m_0}\\
        \tilde E_{a_j}&\neq\tilde E_k+(p_j+\dots+p_1)\omega_d \quad \Rightarrow \quad m_j>0. \label{eq:constraint_m_neq0}
    \end{align}
    This ensures the absence of divergences.
    \item The multiplicity coefficients $c_{m_{r-1},\dots,m_{1}}$ are deduced from the tables of degenerate perturbation theory introduced in Appendix~\ref{app:DPT}.
\end{enumerate}
In the simple case where there is no resonant intermediate state, we obtain $m_j=1$ and the contribution has a trivial multiplicity,~$c_{1,\dots,1}=1$. However, higher-order processes with resonant steps are very common. Such processes can be interpreted as a combination of lower-order processes, as explained in more detail for the example of the sub-harmonic Rabi model in Sec.~\ref{sec:RabiHigherOrder}.
Note that the coefficients $\corr{c_{m_{r-1},\dots,m_{1}}}$ are zero if the combination of powers $\corr{m_{r-1},\dots,m_{1}}$ does not exist under the constraints from above. This explains, for example, the exclusion of resonant intermediate states in the second-order expansions in Eq.~\eqref{eq:Rabi_order2_XZ} and Eq.~\eqref{eq:Stark_order2_XZ}.






\subsection{Leading-order monochromatic Hamiltonian with two resonant states}
\label{sec:Leading_mono_2states}

To exemplify the developed framework, we apply it to a simple, but commonly encountered model with a monochromatic drive, $V_{p}=0$ for $|p|> 1$, and with only two relevant resonant states~$D=\{\ket{0}, \ket{1}\}$. 
%
In this case, the leading order of the coupling rate and Stark shift take a simpler form.
With a single drive harmonics, the system can absorb or emit only one or zero photons at each step and~$p_j=-1,0,1$.
Since the transition between~$\ket{0}$ and~$\ket{1}$ requires the absorption of~$n_1$ photons, the coupling rate vanishes at order $r<n_1$, and the $n_1$-th order coupling corresponds to~$n_1$ successive single photon absorption processes with~$p_j=1$.
Therefore, no intermediate state is resonant with~$E_0$ or~$E_1$ and we only obtain trivial powers and multiplicity coefficients.
This leads to $\Omega_{10}^{(r)}=0 \quad\text{for}\quad r<n_1$ and the leading-order coupling rate
\begin{widetext}
    \begin{align}
        \Omega_{10}^{(n_1)}=\sum_{\ket{a_1},\dots,\ket{a_{n_1-1}}}\frac{V_{1,1a_{n_1-1}}V_{1,a_{n_1-1}a_{n_1-2}}\dots V_{1,a_2a_1}V_{1,a_1 0}}{(\tilde E_0+(n_1-1)\omega_d-\tilde E_{a_{n_1-1}})\dots(\tilde E_0+2\omega_d-\tilde E_{a_2})(\tilde E_0+\omega_d-\tilde E_{a_1})}. \label{eq:lowestRabiMonochrom}
    \end{align}
\end{widetext}
We note that the sum over $\ket{a_1},\dots,\ket{a_{n_1-1}}$ runs in principle over all states of the system. However, in practical applications only a few levels remain relevant and one may ignore those for which the effective coupling to all other resonant states is small enough. Without a quantitative metric, the cut-off is typically obtained heuristically, 
as we shall employ in the following section on applications as well.

In contrast, the Stark shift is already non-zero at order~$r_H=2$ for any~$n_1$
and its leading contribution is given by
\begin{align}\label{eq:effective_energy_order2_monochrom}
    \delta_k^{(2)} =&\, \sum_{l\neq k} \frac{|V_{0,lk}|^2}{\tilde E_k-\tilde E_l}  \nonumber \\
    &+ 
    \sum_l \left(
    \frac{|V_{1,lk}|^2}{\tilde E_k+\omega_d-\tilde E_l}
    + \frac{|V_{-1,lk}|^2}{\tilde E_k-\omega_d-\tilde E_l}
    \right).
\end{align}
The first term is the DC-Stark shift due to the static perturbation~$V_0$.
The last two terms are the AC-Stark shifts associated with the virtual absorption or emission of a drive photon.
Such expression can then be used to determine also in the more general case the resonance frequency, as explained in Eq.~\eqref{eq:ResonanceCondition}.

\subsection{Fast oscillations and leakage}\label{sec:eff_dynamics_pertb}

As illustrated in Fig.~\ref{Fig:2} and by the example in Eq.~\eqref{eq:cl_XY},  the computation of the system's state evolution requires, in addition to $\H_\eff$, knowledge of the transformation $W$ between the full and the reduced Sambe space. In Appendix~\ref{app:DPT} we show how the matrix elements of $W$ can be evaluated alongside the effective Hamiltonian. 
At zeroth order, we have $\corr{W^{(0)}}=P$ such that $\corr{W^{(0)}}\kket{k,n_k}=\kket{k,n_k}$.
At higher order $r\geq 1$, $W$ maps the state $\kket{k,n_k}$ on the other states in the Sambe space according to
\begin{multline}\label{eq:W_mat_elmt}
    \bbra{l, p}\corr{W^{(r)}}\kket{k, n_k}
        =\\
        \sum_{\substack{ p_1,\dots,p_r \\ m_1,\dots,m_{r} \\ |a_1\rangle, \dots,|a_{r-1}\rangle}} c^W_{m_{r},\dots,m_{1}}
        \prod_{j=1}^{r}\frac{V_{p_j,a_j a_{j-1}}}{(\tilde E_k + \sum_{j'=1}^j p_{j'}\omega_d - \tilde E_{a_j})^{m_j}},
\end{multline}
where the sum satisfies the rules:  
\begin{enumerate}
    \item The total number of absorbed photons must satisfy 
    \begin{equation}
        p_1+\dots+p_r = p-n_k.
    \end{equation}
    \item The $r-1$ virtual states $\ket{a_1},\dots,\ket{a_{r-1}}$ are any states of the system, $\ket{a_{0}}\equiv\ket{k}$ and $\ket{a_{r}}\equiv\ket{l}$.
    \item The exponents $m_1,\dots,m_{r}\in \mathbb{N}$ satisfy 
    \begin{equation}
        m_1 + \dots + m_{r} = r,
    \end{equation}
    and they are zero for resonant intermediate states (if any) and non-zero otherwise:
    \begin{align}
        \tilde E_{a_j}&=\tilde E_k+(p_j+\dots+p_1)\omega_d \quad \Rightarrow \quad m_j=0, \\
        \tilde E_{a_j}&\neq\tilde E_k+(p_j+\dots+p_1)\omega_d \quad \Rightarrow \quad m_j>0.
    \end{align}
    \item The multiplicity coefficients $c^W_{m_r,\dots,m_1}$ are different from the coefficients for effective Hamiltonian and are listed in the corresponding tables in Appendix~\ref{app:DPT}.
\end{enumerate}

Given the matrix elements in Eq.~\eqref{eq:W_mat_elmt}, we can use Eq.~\eqref{eq:evol_amplitude_allOrders} to derive the time-evolution of the system wavefunction at a given order in the expansion of $W$.
In most cases, however, the first order expansion of $W=P+\corr{W^{(1)}}+\mathcal{O}(\lambda^2)$ already provides a sufficiently accurate estimate for the fast oscillations and for leakage. 
Within this approximation, we obtain at zeroth order
\begin{align}\label{eq:evol_orderW0}
    c_l^{(0)}(t) = \begin{cases}
        e^{-i(E_0+n_l\corr{\omega_d}) t}\sum_{k\in D} U_{lk}^{\eff}(t)c_k(0) &\text{for }l\in D \\
        0 &\text{for }l\notin D, \\
    \end{cases}
\end{align}
where $U_{lk}^\eff(t)=\bbra{l,n_l}e^{-i\H_\eff t}\kket{k,n_k}$ is the time-evolution operator associated with the effective Hamiltonian.
At first order, we obtain 
\begin{align}
    c_l^{(1)}(t)=&
    \sum_{k,k'\in D}\sum_{p\neq n_l-n_{k'}}
    \frac{e^{-i(E_0+(p+n_{k'})\omega_d)t}V_{p,lk'}U_{k'k}^\eff(t)c_k(0)}{ \tilde E_{k'}+p\omega_d-\tilde E_{l}}\nonumber\\
    +& \sum_{k,k'\in D}\sum_{p\neq n_{k'}-n_k}
    \frac{e^{-i(E_0+n_l\omega_d)t}U_{lk'}^\eff(t)V_{p,k'k}c_k(0)}{ \tilde E_{k'}-p\omega_d-\tilde E_{k}},
\end{align}
for $l\in D$ and 
\begin{align}
    c_l^{(1)}(t) = \sum_{k,k'\in D}\sum_{p}
    \frac{e^{-i(E_0+(p+n_{k'})\omega_d)t}V_{p,lk'}U_{k'k}^\eff(t)c_k(0)}{ \tilde E_{k'}+p\omega_d-\tilde E_{l}}
\end{align}
for $l\notin D$. Note  that the first-order correction usually leads to a non-normalized state. However, in practice it is often enough to evaluate the normalized wavefunction
\begin{align}\label{eq:evol_first_order}
    c_l(t) = \frac{c_l^{(0)}(t)+c_l^{(1)}(t)}{\left(\sum_m |c_m^{(0)}(t)+c_m^{(1)}(t)|^2\right)^{1/2}}
\end{align}
to obtain a sufficiently accurate prediction of the fast oscillations and of leakage. 

The analytical or numerical evaluation of the expansions of the effective Hamiltonian~\eqref{eq:Heff_matEl_general}, transformation $W$~\eqref{eq:W_mat_elmt}, and wavefunction evolution~\eqref{eq:evol_amplitude_allOrders} can be easily automatized using a formal calculus programming language. We developed such a sample code in Python~\cite{ZenodoCode}.

\corr{\subsection{Comparison to other methods}}

\corr{Before we proceed with an illustration of this method for two basic examples, let us briefly compare it to various related approaches for deriving the effective dynamics of Floquet systems. Different perturbation theory formulations can be used in the Sambe space to derive a static effective Hamiltonian, such as the Schrieffer-Wolff~\cite{romhanyi_subharmonic_2015}, van Vleck~\cite{eckardt_high-frequency_2015}, or Brillouin-Wigner~\cite{mikami_brillouin-wigner_2016} approach.
These methods can lead to a different effective Hamiltonian~$\H_\eff$ and transformation~$W$ satisfying Eq.~\eqref{eq:Heff_definition}, since unitary transformations within the degenerate subspace are allowed~\cite{klein1974DegeneratePerturbationTheory,suzuki_degenerate_1983}.
Compared to these methods, the degenerate perturbation theory introduced in Ref.~\cite{soliverez_effective_1969} and used in the current approach has the advantage of expressing~$\H_\eff$ and~$W$ at arbitrary order solely in terms of the projection~$P$, the perturbation~$\V$, and the resolvent operator~$R$.
This enables us to obtain the closed expressions in Eq.~\eqref{eq:Heff_matEl_general} and Eq.~\eqref{eq:W_mat_elmt} at order~$r$, without the computation of either the previous~$r-1$ orders or intricate nested commutators, which become hard to track at higher orders. 
}
\corr{
Our formulation also enables the diagrammatic representation of Fig.~\ref{fig:diagr_order2_general}, which, to our knowledge, has not been introduced before. This representation is particularly useful for identifying in which cases multi-photon transitions are allowed, depending on the selection rules of the drive operator, as illustrated for the Rabi model below.
}

\corr{
The Floquet-Magnus expansion is another method to derive an effective static Hamiltonian without using the Sambe space~\cite{magnus_exponential_1954,casas2001Floquet,rahav2003Effective,goldman_periodically_2014,dey2025ErrorBoundsFloquetMagnus} and a generalization thereof has been recently described in Ref.~\cite{venkatraman2022Static}. These methods also require iterative calculations of nested commutators, which become impractical at high orders. Moreover, they are based on a high-frequency expansion, which is valid when the driving frequency~$\omega_d$ is large compared to the system's typical frequency transitions. If this is not the case, but energy levels are of the form $E_k=E_0+ n_k\omega_d+\epsilon_k$, the high-frequency expansion can still be applied by moving first to a rotating frame and if $\epsilon_k\ll\omega_d$, i.e., the spectrum is weakly anharmonic.
%
%
This is usually not the case when considering a subharmonic driving for an arbitrary anharmonic system, as illustrated in Fig.~\ref{fig:levels_notations}, where~$\omega_d$ can be of the same order as the frequency detuning~$\epsilon_l$ of the non-resonant states~$\ket{l}\notin D$.
}
\corr{
A similar limitation applies to the method discussed in Ref.~\cite{xiao2024}, which is also specifically developed for weakly nonlinear oscillators, such as transmon qubits, which can be expressed in terms of bosonic annihilation and creation operators. This approach is thus not readily applicable for generic few-level systems or the fluxonium discussed in Sec.~\ref{sec:fluxonium} below, where the driving frequency is much smaller than the anharmonicity. Our approach distinguishes the quasi-resonant and non-resonant states, and treats  the detuning~$\epsilon_l$ of non-resonant states~$\ket{l}\notin D$ non-perturbatively.
This is why differences between energies of the undriven system appear in the general expressions like Eq.~\eqref{eq:Heff_matEl_general}, and not only the driving frequency~$\omega_d$ as obtained in high-frequency expansions.
}

\corr{
Importantly, the different methods mentioned above focus on the effective Hamiltonian and do not provide a systematic perturbative expansion of the time-dependent wavefunction and its fast oscillations, see Eq.~\eqref{eq:evol_amplitude_allOrders} and Eq.~\eqref{eq:W_mat_elmt}.
As shown in the next sections, resolving these fast oscillations is essential for high-fidelity quantum control applications.
}

\section{Application: Rabi model}\label{sec:RabiModel}

In this section, we illustrate the application of our general framework to the case of a subharmonically driven Rabi model with Hamiltonian
\begin{equation}
    H_{R}(t) = \frac{\omega_{01}}{2}\sigma_z + 2\Omega_x\cos(\omega_d t)\sigma_x
\end{equation}
and a driving frequency that satisfies $\omega_d\simeq \omega_{01}/n_1$. This model is a special case of the $XZ$ model with $\Omega_{z}=0$ and in the notation of Sec.~\ref{sec:RF-P} we have~$V_1=V_{-1}=\Omega_x\sigma_x$. Note that the subharmonic Rabi model has been investigated previously in the literature, but mainly in the presence of damping. In this case, the steady state population can be derived as a function of the driving frequency in a continued fraction expansion~\cite{chakmakjian_observation_1988,ruyten_subharmonic_1989,koch_subharmonic_1989} and exhibits local maxima at the subharmonic resonance frequencies.
In the absence of damping, the leading order coupling rate and Stark shift are obtained in~\cite{shirley_solution_1965}.


\subsection{Leading order Rabi frequency}
The Rabi model has the specific property $V_{1,00}=V_{1,11}=0$, such that only virtual transitions that alternate between the states ~$\ket{0}$ and~$\ket{1}$ take place. 
Therefore, the leading order coupling rate given in Eq.~\eqref{eq:lowestRabiMonochrom} vanishes for even~$n_1$. For odd $n_1=2q+1$, it gives
\begin{align}
    \Omega_{10}^{(n_1)}=\frac{V_{1,10}V_{1,01}\dots V_{1,10}}{2q\omega_{d}(-2\omega_{d})(2q-2)\omega_d\dots(2\omega_d)(-2q)\omega_d},
\end{align}
which corresponds to $q+1$ transitions from~$\ket{0}$ to~$\ket{1}$ and ~$q$ transitions back from~$\ket{1}$ to~$\ket{0}$.  Each transition is accompanied by the absorption of one photon of energy $\omega_{d}$.
This process is represented diagrammatically in Fig.~\ref{fig:Sub-harmonic_Rabi}(a) for $n_1=3$ and in Fig~\ref{fig:Sub-harmonic_Rabi}(b) for $n_1=5$. The energies involved in the denominator are represented by the red arrows (the orientation encoding the sign).
Note that the naive RWA amounts to considering~$V_{1,01}=0$. This approximation ignores the photon absorption accompanied by a virtual transition from $\ket{1}$ to $\ket{0}$ and leads to a vanishing coupling rate for $n_1\geq 2$.

For an odd resonance, the coupling rate can be written in a more compact form as
\begin{align} \label{eq:SR_prefactor}
    \Omega_{10} &=  \frac{(-1)^q}{2^{2q}(q!)^2}\frac{\Omega_x^{2q+1}}{\omega_{d}^{2q}}+ \O(\lambda^{n_1+1}), 
\end{align}
showing the expected scaling~$\sim \Omega_x^{n_1}/\omega_{d}^{n_1-1}$. Here one has to keep in mind that $\omega_d=\omega_{01}/n_1+\O(\lambda)$, and when expressed in terms of the level splitting $\omega_{01}$ there is an additional numerical prefactor $\sim n_1^{2q}$. 
By using the Stirling formula to approximate
$q!\sim \sqrt{2\pi q}(q/e)^q$ as well as $(2q+1)^{2q}\sim e(2q)^{2q}$, where $e\approx2.718$ is the Euler number, we obtain 
\begin{equation}
     \Omega_{10}\sim \frac{(-1)^{\frac{n_1-1}{2}}}{\pi (n_1-1)}\left(\frac{e\Omega_x}{\omega_{01}}\right)^{n_1} \omega_{01} + \O(\lambda^{n_1+1})
\end{equation}
for large odd $n_1$.
As such, we expect to get experimentally accessible Rabi oscillations even at very large subharmonic orders when the driving strength approaches a value of~$\Omega_x\approx \omega_{01}/e\simeq 0.37\omega_{01}$.

\subsection{Leading order Stark shifts}
In contrast to the Rabi frequency, the Stark shift is non-zero at order $r=2$ for any~$n_1$. The two processes contributing to the Stark shift
are similar to the ones depicted in Fig.~\ref{fig:diagr_order2_general}(c) for the $XZ$ model and lead accordingly to a second-order contribution 
\begin{equation}\label{eq:RabiStark} 
    \delta_{0}^{(2)} = -\frac{\Omega_x^2}{(n_1+1)\omega_{d}} -\frac{\Omega_x^2}{(n_1-1)\omega_{d}}
     \simeq -\frac{2n_1^2\Omega_x^2 }{(n_1^2-1)\omega_{01}}
\end{equation}
and $\delta_{1}^{(2)} = - \delta_{0}^{(2)}$ for any $n_1\neq 1$.  By adding the first-order contributions, $\delta^{(1)}_{1}=\epsilon_1=\omega_{01}-n_1\omega_d$ and $\delta_0^{(1)}=0$, the dominant correction for the actual resonant driving frequency $\omega^{\rm res}_d$ can be obtained from the solution of 
\begin{equation}\label{eq:RabiCorrection}
    n_1\omega_d^{\rm res} = \omega_{01} + \frac{2\Omega_x^2}{(n_1+1)\omega_d^{\rm res}}+\frac{2\Omega_x^2}{(n_1-1)\omega_d^{\rm res}}.
\end{equation}
Note that resulting correction is almost independent of the subharmonic order $n_1>1$.


\subsection{Higher order processes}\label{sec:RabiHigherOrder}
In contrast to the two-photon $XZ$ model studied above, the second-order correction in Eq.~\eqref{eq:RabiCorrection} is in general not accurate enough to induce resonant oscillations,  since for any $n_1\geq3$, the leading order coupling rate $\Omega_{10}^{(n_1)}$ scales with a higher power of the drive strength. Similarly, when increasing the  drive strength to $\Omega_x\approx \omega_{01}/e$ to boost the $n_1$-photon transition, processes of order $n_1+2$ lead to significant corrections to the Rabi-frequency and must be taken into account to obtain a sufficiently high accuracy. 

\begin{figure}
    \centering
    \includegraphics[width=\columnwidth]{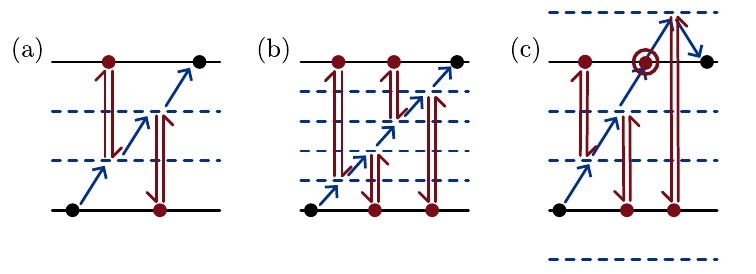}\ \ 
    \caption{{\bf Sub-harmonic Rabi model with monochromatic drive}. (a,b) Diagrammatic representation of the lowest order coupling rate for $n_1=3$ and $n_1=5$, following the same notation as explained in Fig.~\ref{fig:diagr_order2_general}.
    (c) Diagrammatic representation of one type of process for the fifth-order coupling rate at $n_1=3$.
    The encircled red dot is a resonant step that would naively induce a diverging process and must be assigned a nontrivial multiplicity coefficient.
    }
\label{fig:Sub-harmonic_Rabi}
\end{figure}

To illustrate the evaluation of some of the non-trivial higher-order processes in the Rabi model, we consider the diagram shown in Fig.~\ref{fig:Sub-harmonic_Rabi}(c) as a concrete example. It represents a contribution to the coupling rate at order $r=5$ in the case of an $n_1=3$ photon resonance. As shown in this diagram, after absorbing three photons, the system reaches a resonant intermediate state, before undergoing two additional off-resonant transitions. 
This resonant virtual state is encircled in red and would naively result in a vanishing energy denominator (no corresponding red arrow).
In this case, according to the rules summarized in Sec.~\ref{sec:arbitrary_order_Heff}, we must set~$m_3=0$, and the other powers $m_j$ are non-zero and must add up to $r-1=4$. 
The possible configurations are $(m_1,m_2,m_3,m_4)=(2,1,0,1), (1,2,0,1), (1,1,0,2)$.
According to Table~\ref{table:DPTcoeff_order5} in Appendix~\ref{app:DPT}, all these combinations have a multiplicity coefficient of $c_{m_1,m_2,m_3,m_4}=-1/2$.
The energy penalty of the non-resonant steps can be read off from the red arrows and is given by $-2\omega_{d}$, $2\omega_{d}$, and $4\omega_{d}$ for $j=1,2,4$, respectively.
Then, the amplitude of the process of \corr{Fig.~\ref{fig:Sub-harmonic_Rabi}(c)}, which corresponds to a succession of four  absorbed photons followed by one emission, is
\begin{align}
    -\frac{1}{2}\left(
    \frac{1}{(-2)^2\cdot2\cdot4}
    +
    \frac{1}{-2\cdot2^2\cdot4}
    +
    \frac{1}{-2\cdot2\cdot4^2}
    \right)
    \frac{\Omega_x^5}{\omega_{d}^4}
    .
\end{align}
The total fifth-order coupling rate also contains the processes where the photons are absorbed and emitted in a different order.
This shows that it becomes rapidly tedious to keep track of all the processes involved at higher orders.
However, the evaluation of all possible contributions can be easily automated with a formal calculus programming language.
We developed such a Python code to obtain higher-order analytical formulas for such models~\cite{ZenodoCode}.


\subsection{Effective dynamics}
\label{sec:rabi_effdyn}

\begin{figure}
    \centering
    \includegraphics[width=.99\linewidth]{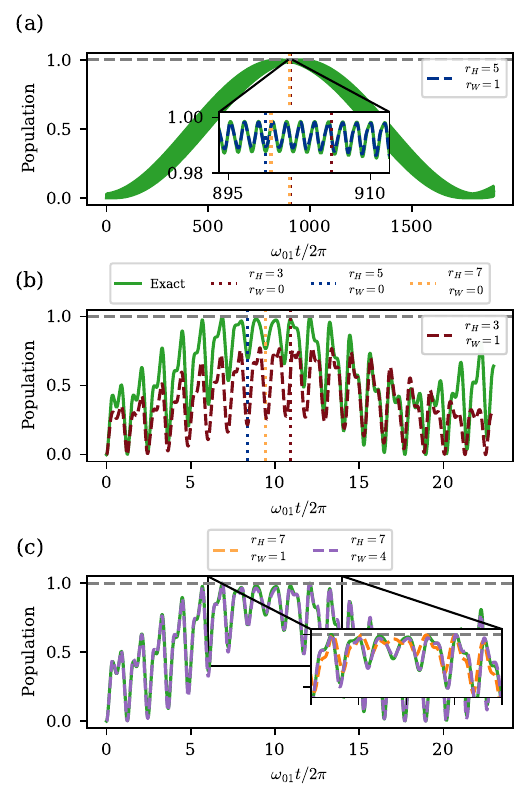}
    \caption{
    \textbf{Dynamics of the $n_1=3$ sub-harmonic Rabi model.}   
    The green solid lines show the full numerical solutions for the population of state $\ket{1}$ for (a) $\Omega_x/\omega_{01}=0.05$ and (b,c) $\Omega_x/\omega_{01}=0.25$.
    The vertical dotted lines in red, blue and yellow indicate the half Rabi period $t_\pi=\pi/\Omega_R^{[r_{H}]}$ predicted at orders $r_{H}=3,5,7$, respectively. 
    All other lines show the predicted state evolution obtained from Eq.~\eqref{eq:evol_amplitude_allOrders} from the computation of the effective Hamiltonian and the operator $W$ at the orders indicated in the legend. 
    }
    \label{fig:evol_rabi_model}
\end{figure}

To demonstrate the application of our effective theory for evaluating the full dynamics of the subharmonically driven Rabi model, we choose $n_1=3$ and compare in Fig.~\ref{fig:evol_rabi_model} the results from perturbation theory with an exact numerical integration of the Schr\"odinger equation. More precisely, in this plot, it is assumed that the system is initially prepared in state $|0\rangle$ and we plot the population of the bare eigenstate $|1\rangle$ as a function of time and for two different values of the driving strength, $\Omega_x/\omega_{01}=0.05$ and $\Omega_x/\omega_{01}=0.25$. For a given value of~$\Omega_x$, we set the drive frequency~$\omega_d$ to a numerically identified value, which optimizes the transferred population in the full simulation. 
This frequency~$\omega_d$ is  held fixed and also used in the perturbative calculations, independently of the order of perturbation theory. The prediction of the resonance frequency is studied more systematically in the next section for the example of a driven fluxonium.

The evolution of the population $p_1(t)=|\langle 1|\psi(t)\rangle|^2$ in Fig.~\ref{fig:evol_rabi_model} shows almost complete Rabi oscillations with the additional fast oscillations discussed in Sec.~\ref{sec:eff_dynamics} on top.
When we compute the effective Hamiltonian at a given order $r_H$ and approximate the operator $W$ at zeroth order, the resulting predicted dynamics in  Eq.~\eqref{eq:evol_orderW0} corresponds to simple Rabi oscillations at a Rabi frequency
\begin{equation}
    \Omega_R^{[r_{H}]} = \sqrt{\left(\delta_1^{[r_{H}]}-\delta_0^{[r_{H}]}\right)^2+4\left|\Omega_{10}^{[r_{H}]}\right|^2}.
\end{equation}
This frequency accounts for the effect of a residual Stark shift between the two states, which varies for different~$r_{H}$.
The vertical dotted lines in Fig.~\ref{fig:evol_rabi_model} correspond to the half Rabi period $t_\pi=\pi/(\Omega_R^{[r_{H}]})$ predicted by the effective Hamiltonian computed at orders $r_{H}=3,5,7$.
%
%
For small driving strength, $\Omega_x/\omega_{01}=0.05$, the leading third order contribution already accurately predicts  the period of Rabi oscillations, with insignificant corrections at higher orders [see Fig.~\ref{fig:evol_rabi_model}(a)]. 
\corr{
At order $r_H=3$, the fast oscillations of the populations are then predicted by the computation of $W$ at first order with an accuracy of roughly 5\%. By increasing the order in the expansion of the Hamiltonian to $r_H=5$ ($r_H=7$), the maximal deviation from the exact evolution reduces to 0.0015 ($<$0.0009).
}
As seen in the inset of Fig.~\ref{fig:evol_rabi_model}(a), the calculation of these fast oscillations is required to predict the exact time of maximal population transfer \corr{at this sub-percent level of accuracy.}

At a larger amplitude $\Omega_x/\omega_{01}=0.25$, as shown in Fig.~\ref{fig:evol_rabi_model}(b,c), the higher order corrections for $\Omega_R^{[r_{H}]}$ are very relevant. 
Specifically, in this example the effective Hamiltonian computed at order $r_H=3$ and fast oscillations computed at order $r_W=1$ predict a maximum population of about $75\%$ [red dashed line in Fig.~\ref{fig:evol_rabi_model}(b)], due to an incorrect prediction of the Stark shift.
Computing the effective Hamiltonian at order $r_H=7$ gives an accurate prediction of the Stark shift, with complete Rabi oscillations in the Sambe representation.
The amplitudes of the fast oscillating components of the actual state evolution are still captured to a good approximation by the first order correction of $W$ [yellow dashed line in Fig.~\ref{fig:evol_rabi_model}(c)].
However, to reach full agreement with the exact system evolution, the computation of $W$ up to order $r_W=4$ is necessary [purple dashed line in Fig.~\ref{fig:evol_rabi_model}(c)]. At this order, perturbation theory can be used to predict the times of maximal population transfer at subpercent precision.

\section{Application: Fluxonium}
\label{sec:fluxonium}

In this section, we will apply the degenerate Floquet perturbation theory to a particular test case to see its scope and limitations for a realistic experimental setup. The fluxonium qubit is a prime example for this, due to the possibility to compare our analysis to recent experimental investigations of sub-harmonic control in such systems~\cite{schirk_subharmonic_2025}. We remark that sub-harmonic driving has already been investigated in other superconducting circuits, for example, for transmons~\cite{xia_fast_2025}. However, the fluxonium presents a relevant case where the previously employed tools, such as the RWA, fail. This is mainly due to their significant anharmonicity, low drive frequency and failure of a two-level approximation, as we demonstrate in Appendix~\ref{app:transmon}. In contrast, the method of degenerate perturbation theory outlined in this manuscript predicts accurately the Rabi rate as well as a small amplitude-dependent shift of the drive frequency even when driving the system at comparably large amplitudes.

\subsection{Setup}

As shown in~\cref{fig:levels_notations}(a), a fluxonium qubit consists of a single Josephson junction with Josephson energy $E_\mathrm{J}$, shunted by a pair of capacitive pads with a charging energy $E_\mathrm{C}$, and an inductance with inductive energy $E_\mathrm{L}$. The Hamiltonian of the fluxonium circuit at its biased sweetspot reads~\cite{you_circuit_2019,didier_ac_2019}
\begin{align}
    H &= H_q + H_d(t),
\end{align}
where 
\begin{align}
    H_q &= 4E_\mathrm{C} n^2 + E_\mathrm{J} \cos \varphi + \frac{E_\mathrm{L}}{2} \varphi^2 = \sum_{k}E_k \ketbra{k}{k}
\end{align}
is the bare fluxonium Hamiltonian. Here $n$ and $\varphi$  denote the canonical conjugate phase and charge operators, which obey $[\varphi,n]=i$. The driving term, induced by a periodically modulated external magnetic flux\corr{~\cite{schirk_subharmonic_2025}}, is given by
\begin{align}
    H_d(t) &= - E_L A\cos(\omega_d t) \varphi = \bar{V}_{1}e^{-i\omega_dt} + \bar{V}_{-1}e^{i\omega_dt},
\end{align}
%
%
where $\omega_d$ is the frequency and $A$ the dimensionless amplitude of the modulation. 
Explicitly, when expressed in terms of the fluxonium eigenstates $|k\rangle$, we obtain the harmonics 
\begin{equation}
 \bar{V}_{1}=\bar{V}^\dagger_{-1}=-\frac{A E_\mathrm{L}}{2} \sum_{i,k}\bra{i}\varphi\ket{k}\ketbra{i}{k}=\sum_{i,k}V_{ik}\ketbra{i}{k}   
\end{equation}
and $\bar V_{p}=0$ for $|p|\neq 1$. We emphasize that due to the symmetry of the fluxonium potential, the matrix elements $V_{1,ik}\equiv V_{ik}$ are non-zero only for $|i-k|=2m+1, m\in \mathbbm{N}$.
\corr{Note that charge driving has the same structure as flux driving in $\bar{V}_{\pm 1}$, only up to matrix elements in $V_{ik}$ with different amplitudes.}

For the following discussion, we focus on the case $n_1=3$, i.e., we drive the fluxonium at about a third of the transition frequency between the lowest two eigenstates $|0\rangle$ and $|1\rangle$. 
This corresponds to the decomposition 
\begin{equation}
E_1 - E_0 = 3 \omega_d + \epsilon_1
\end{equation}
in~\cref{eq:energy_decomposition}. We introduce the detuning $\epsilon$ from three-photon resonance, where $\epsilon\equiv 3\omega_d-(E_1-E_0) =-\epsilon_1 \ll \omega_d$. We assume that there are no other accidental  resonances, $|\epsilon_l| \gg |\epsilon|,\; \forall l\neq1$, and restrict the degenerate subspace to $D=\{ |0\rangle, |1\rangle\}$. 
Based on the maximal population transfer determined by optimizing exact time-evolutions, as later shown in ~\cref{sec:numerical_verification}, we find that the total Hilbert space can be effectively truncated to states \( |a\rangle \) for \( a \leq 4 \). 
It is supported by the fact that including higher levels does not change the resonant fidelity significantly. 
Thus, to accurately capture the dynamics of the system, we adopt this truncation in the following setting.



\subsection{Leading order Hamiltonian}
We apply the general formalism from above to evaluate the effective Hamiltonian for the subharmonically driven fluxonium,
\begin{equation}
    \label{eq:eff_fluxonium_hamiltonian}
    \H_{\eff} = \begin{pmatrix}
        \delta_{0} & \Omega_{10}^* \\
        \Omega_{10} & \delta_{1}
    \end{pmatrix},
\end{equation}
with $\Omega_{10}=\sum_{r=1}^{r_{H}} \Omega_{10}^{(r)}$ and $\delta_{k}=\sum_{r=1}^{r_{H}} \delta_{k}^{(r)}$.
We justify that due to the non-harmonic feature of the fluxonium, states with different parity have non-zero dipole moments, $V_{ij}\neq0, \forall  |i-j|=2k+1,k\in\mathbb{N}$.
Since $\Omega_{10}^{(1)}=\Omega_{10}^{(2)}=0$, the leading contribution to the effective coupling is 
\begin{equation}
    \begin{aligned}
    &\Omega_{10}^{(3)}=
    -\frac{V_{10}V_{01}V_{10}}{4\omega^2_\mathrm{d}}+\frac{V_{10}V_{03}V_{30}}{2\omega_\mathrm{d}(\tilde{E}_0-\tilde{E}_3+\omega_\mathrm{d})}\\
    &-\frac{V_{12}V_{21}V_{10}}{2\omega_d(\tilde{E}_0-\tilde{E}_2+2\omega_d)}+\frac{V_{12}V_{23}V_{30}}{(\tilde{E}_0-\tilde{E}_2+2\omega_d)(\tilde{E}_0-\tilde{E}_3+2\omega_d)}\\
    &-\frac{V_{14}V_{41}V_{10}}{2\omega_d(\tilde{E}_0-\tilde{E}_4+2\omega_d)}+\frac{V_{14}V_{43}V_{30}}{(\tilde{E}_0-\tilde{E}_3+2\omega_d)(\tilde{E}_0-\tilde{E}_4+2\omega_d)}.
\end{aligned}
\label{eq:f-p_rabi}
\end{equation}
%
%
%
Up to the same order,  we obtain for the relevant energy splitting $\Delta=\delta_1-\delta_0=\sum_r \Delta^{(r)}$ the first order contribution $\Delta^{(1)}=\epsilon_1=-\epsilon$ and 
\begin{align}
&\begin{aligned}
    \Delta^{(2)}=&\frac{3|V_{01}|^2}{2\omega_d}-\frac{|V_{03}|^2}{\omega_\mathrm{d}+\tilde{E}_0-\tilde{E}_3}-\frac{|V_{03}|^2}{-\omega_\mathrm{d}+\tilde{E}_0-\tilde{E}_3}\\
    &+\frac{|V_{12}|^2}{2\omega_d+\tilde{E}_0-\tilde{E}_2}+\frac{|V_{12}|^2}{4\omega_d+\tilde{E}_0-\tilde{E}_2}\\
    &+\frac{|V_{14}|^2}{2\omega_d+\tilde{E}_0-\tilde{E}_4}+\frac{|V_{14}|^2}{4\omega_d+\tilde{E}_0-\tilde{E}_4}, \label{eq:f-p_shifts_2}
\end{aligned}
    \\
&\begin{aligned}
    \Delta^{(3)} = &- \frac{5|V_{01}|^2\epsilon}{8\omega_d^2} \\
    &-\frac{|V_{12}|^2\epsilon}{(4\omega_d+\tilde{E}_0-\tilde{E}_2)^2}- \frac{|V_{12}|^2\epsilon}{(2\omega_d+\tilde{E}_0-\tilde{E}_2)^2}\\
    &-\frac{|V_{14}|^2\epsilon}{(4\omega_d+\tilde{E}_0-\tilde{E}_4)^2}-\frac{|V_{14}|^2\epsilon}{(2\omega_d+\tilde{E}_0-\tilde{E}_4)^2}.\label{eq:f-p_shifts_3}
\end{aligned}
\end{align}
Note that both odd and even orders contribute to the Stark shifts, which is a consequence of the non-parity-changing term $V_{0,11}=\epsilon_1$. Beyond the third order, this static perturbation contributes to the expressions for $\Omega_{10}^{(4)}, \Omega_{10}^{(5)}, \dots$ as well.


\subsection{Numerical verification} \label{sec:numerical_verification}
To test the accuracy of Floquet perturbation theory for realistic quantum circuits, we consider a set of parameters close to the experimental values in Ref.~\cite{schirk_subharmonic_2025}. Specifically, we set $E_\mathrm{J}/h=1.69$\,GHz, $E_\mathrm{L}/h=1.07$\,GHz and $E_\mathrm{C}/h=0.68$\,GHz. From a numerical diagonalization of $H_q$ we then obtain $ (E_1-E_0)/h=1.33\; \text{GHz},\;(E_2-E_1)/h= 2.15 \; \text{GHz}$ and 
\begin{align}
    &\frac{V_{01}}{h}=\frac{E_L\langle 0|\varphi|1\rangle A}{2h}=4.72 \times \frac{A}{2\pi} \; \text{GHz},\\
    &\frac{V_{12}}{h}=\frac{E_L\langle 1|\varphi|2\rangle A}{2h}=5.28 \times \frac{A}{2\pi} \; \text{GHz}. 
\end{align}
Note that for convergence of the perturbation theory, we restrict ourselves to the regime where $A/2\pi < 4\omega_d/( E_L \langle 0|\varphi|1\rangle) \approx 0.18$.

\begin{figure}
    \centering
    \includegraphics[width=\columnwidth]{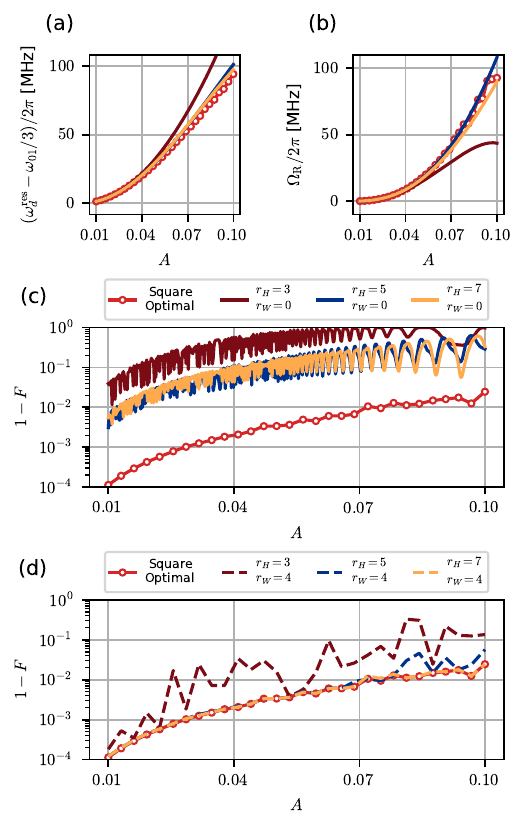}
    \caption{\textbf{Convergence of Floquet perturbation theory for 3-photon sub-harmonically driven fluxoniums.}
    The resonant frequency detuning {\bf (a)} and Rabi rates  {\bf (b)} for driving Rabi oscillations with sub-harmonic processes are numerically determined (red dots). The predictions from degenerate Floquet-perturbation theory at finite orders (solid colors) show convergence towards the numerical values.
    {\bf (c)} The fidelity when using a constant pulse to transfer state $\ket{0}$ to $\ket{1}$ with numerically optimized pulse length. Even when numerically optimizing the resonance condition (red), the fidelity decreases for large drives. 
    {\bf (d)}
    Incorporating $W$ at fourth order (dashed) eliminates fast oscillations of the fidelity, seen by consistently tracking the time of maximum population transfer from~\cref{eq:evol_amplitude_allOrders}. 
    All shown fidelities have been found by numerically optimizing with respect to $\omega_d$. This confirms that once convergence to the resonance condition is reached, the dynamics is precisely predicted.}
    \label{fig:comparison}
\end{figure}

To drive high-fidelity Rabi oscillations, our goal is to accurately predict the Rabi frequency $\Omega_{R}$ together with the three-photon detuning  $\epsilon=3\omega_d-(E_1-E_0)$ that achieves $\Delta=0$. 
When increasing the truncation of the system to be $a\leq 4$ in~\cref {fig:comparison}(a) and (b), we compare the values predicted for both quantities at different orders of perturbation theory with the numerically optimized results. These numerical results are obtained for each value of drive amplitude $A$ by extracting $\epsilon$ and $\Omega_{R}$ from optimized values of the drive frequency $\omega_d$ and from the time $t_{\rm op}$ at which the full time evolution first maximizes the population of \(\ket{1}\), i.e., $\Omega_{R}|_{\rm num} =\pi/t_{\rm op}$. Note that in (b) the numerically optimized solution  visibly jumps at certain higher amplitudes. This jump can be attributed to the fact that the optimal pulse length (and with it the Rabi rate) is determined by the time of maximum population transfer. The latter one corresponds to the highest of several peaks of the fast oscillations, which is not necessarily closest to the averaged maximal population, as was also visible in~\cref{fig:evol_rabi_model}(c). When varying the amplitude, the local peaks grow and fall independently, and the attribute of being the ``highest''  will shift from one to another at some amplitude. This instantaneous change in attribution causes the jump in the Rabi rate.


While for very weak driving fields the leading order contributions $\Delta^{(2)}$ and $\Omega_{01}^{(3)}$ provide very accurate predictions for $\epsilon$ as well as the Rabi frequency $\Omega_R$, we observe significant deviations from the exact values already at driving strengths $ 0.05 \lesssim A/(2\pi)\lesssim 0.1$. In this regime, we obtain a substantial improvement by including the $5$th-order and $7$th-order corrections, at which the predicted frequency detuning and coupling rate are almost converged. This comparison demonstrates that, already at rather weak driving strengths, it is necessary to include processes beyond the leading order to obtain qualitatively accurate results. However, we point out that the number of processes contributing to each order grows exponentially. 
Summing up all processes at high orders, therefore, can become prohibitive. 

To check the accuracy of Floquet perturbation theory at different orders, we demonstrate the fidelity of a $\pi$-rotation for various drive amplitudes in~\cref{fig:comparison}(c). The fidelity of the transfer is given by $F=|\langle 1|\psi(t_\pi)\rangle|^2$, where $t_\pi=\pi/\Omega_R$
 is chosen according to the predicted values of the Rabi frequency at different orders of perturbation theory (solid colors) or according to the numerically optimized results, $t_\pi=t_{\rm op}$ (red dots). As above, the drive frequency is determined by the solution of $\Delta(\omega_d)=0$ at the corresponding orders and by numerical optimization, respectively.  
Fig.~\ref{fig:comparison}(c) demonstrates that, given the $r_H=7$th-order theoretical prediction for the resonant frequency and the Rabi rate, we can achieve population transfer fidelities at the level of $99.5\%$ for small drive amplitudes. 


Nonetheless, for the given parameters of the fluxonium, the Floquet perturbation theory approximately converges around $r_H=5$th order, as increasing to $r_H=7$th order leads to insignificant improvements in~\cref{fig:comparison}(c). 
The remaining difference in fidelities between the Floquet perturbation theory and the numerically optimized solution can be attributed to the fast oscillating components of the state populations during the time evolution under the constant drive, which are clearly visible in the fidelity oscillations in~\cref{fig:comparison}(c).
This more complicated evolution means that the optimal transfer no longer occurs at $\Delta=0$ of the effective two-level model. To illustrate this point, we plot in Fig.~\ref{fig:comparison}(d) the transfer infidelity, evaluated at a time $t_\pi=t_{\rm op}^{(r_H,r_W)}$. Here, the optimal time is evaluated by a numerical search in the same way as the numerically optimized results, but using only $\H_\eff$ and $W$ evaluated up to the respective orders. This comparison shows that for the current task, perturbation theory at order $(r_H=7,r_W=4)$ is essentially indistinguishable from the full numerics, while moderate driving strengths deviations in the percent level may still occur at lower orders.
\corr{
In a generic setting, when the comparison with the exact numerics is not accessible, the minimal order required for a desired fidelity is determined from the convergence of the series, i.e. by checking whether increasing the order improves the fidelity up to a chosen tolerance.
}
Note that for increasing drive amplitude, the fidelity of the state transfer decreases again, even for the numerically optimized solution. 
This can be attributed to leakage out of the computational subspace, i.e., a nonvanishing population of state $|2\rangle$.

\section{Conclusion \& Outlook}
\label{sec:conclusion}
In summary, we have presented a general theoretical framework for analyzing the dynamics of periodically driven quantum systems using degenerate Floquet perturbation theory. This approach is specifically suited for the analysis of multi-photon driving schemes and similar quantum control operations, for which a precise knowledge of the effective couplings and frequency shifts and of the exact state evolution is required. We have illustrated the application of this formalism for describing multi-photon Rabi-oscillations in two-level systems and fluxonium qubits, which showed that even higher-order contributions in the effective Hamiltonian and transformation $W$ can have a significant influence. While this perturbation theory is restricted to weakly driven systems, it does not rely on any other approximations, such as a rotating wave or a two-level approximation, and can be evaluated in a systematic manner without any prior knowledge about the system. To assist in the adaptation of this theory for other applications, we introduced an intuitive diagrammatic representation of all the terms in the expansion and made a sample code for an automated numerical and analytical evaluation of the perturbation series available~\cite{ZenodoCode}. 
By making use of such tools, our framework allows us to evaluate multi-photon processes up to arbitrarily high orders, which in practice is eventually limited by the exponential growth of the terms that contribute at each order.

\corr{Our result differentiates from previously utilized methods in the literature \cite{sarkar_subharmonics_2021,xiao2024,magnus_exponential_1954,goldman_periodically_2014, goldman_periodically_2015, bukov_universal_2015,eckardt_colloquium_2017, schweizer_floquet_2019,rodriguez-vega_low-frequency_2021,eckardt_high-frequency_2015,sah_decay-protected_2024} in both presenting an analytic framework, whose $n$th order does not require iterative computations of the ($n$-1)th order, and by being applicable even to systems with low frequency and high anharmonicity. In Appendix \ref{app:transmon}, we demonstrate how our approach extends into parameter regimes where methods such as RWA lose accuracy.}

\begin{figure}
    \centering
    \includegraphics[width=.99\linewidth]{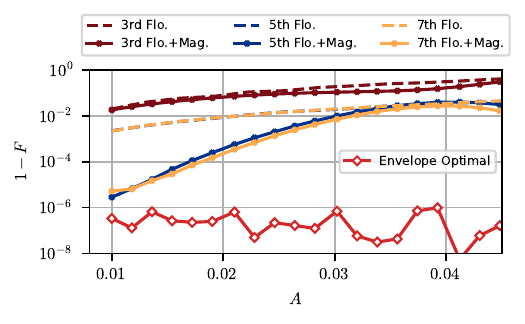}
    \caption{\textbf{Improved fidelities with pulse shaping.} 
    When using an enveloped pulse and numerically optimizing the pulse length, one finds very high fidelities for various drive amplitudes (red dots). 
    The fidelity under a constant drive (dashed colors), as predicted from effective dynamics at finite orders, serves as the reference. 
    The predictions from an adiabatic Magnus-expansions (see Appendix \ref{app:pulse_shaping}) at finite orders (solid colors) are close to the optimized fidelity in small amplitudes. For the given values, the radius of convergence of the Magnus expansion is $A/2\pi \approx 0.05$, explaining the rapid decline in fidelity thereafter. 
    }
    \label{fig:pulse_shaping_fidelity}
\end{figure}

While in the current analysis we have focused exclusively on the time evolution under a square-like pulse of constant amplitude, the effective parameters predicted from this analysis can serve as input for other control schemes to design further optimized and experimentally more relevant pulse shapes.  As an outlook and illustrative example, we discuss in\corr{~\cref{app:pulse_shaping}} the derivation of an optimal shape for the drive amplitude in form of a flat-top Gaussian. This is commonly used in experiments to avoid leakage out of the qubit subspace 
by quasi-adiabatically switching on and off the drive field. 
As demonstrated by the results summarized in~\cref{fig:pulse_shaping_fidelity}, this optimization procedure---based solely on predictions from perturbation theory---allows us to considerably improve the Rabi fidelities obtained under the same assumptions for a simple square pulse in~\cref{fig:comparison}(c). 

In the future, it will be interesting to extend this proof-of-concept example and explore more general applications of high-order perturbation theory for quantum control~\cite{werninghaus_leakage_2021, glaser_sensitivity-adapted_2024}. In particular, our theoretical framework developed here can be readily applied to the control of 0-$\pi$-qubits \cite{gyenis_experimental_2021} or of two-qubit gates \cite{ding_high-fidelity_2023,rosenfeld_high-fidelity_2024, heunisch_scalable_2025}, by making use of higher-order transitions that have not been investigated systematically in such scenarios before. An extension of our theory to multi-tone drives~\cite{ho1983SemiclassicalManymodeFloquet} would further enable the modeling of control schemes in setups where drives with incommensurate frequencies are applied  \cite{gyorgy_electrically_2022, john_bichromatic_2024,gyorgy_limitations_2025}.

\vspace{20pt}
\section*{Acknowledgements}
We thank Lukas Schamriss, Federico Roy, \corr{Daniel Burgarth}, Lukas Heunisch, Michael Hartmann, and Andras Palyi for insightful discussions and helpful comments. This work received financial support from the German Federal Ministry of Education and Research via the funding program under contract number 13N16188 (MUNIQC-SC) and under contract number 13N15680 (GeQCoS) as well as by the European Union by the EU Flagship on Quantum Technology HORIZON-CL4-2022-QUANTUM-01-SGA project 101113946 OpenSuperQPlus100. The research is part of the Munich Quantum Valley, which is supported by the Bavarian state government with funds from the Hightech Agenda Bayern Plus.

\bibliography{references.bib}

@dataset{ZenodoCode,
	title = {Theory of Multi-photon Processes for Applications in Quantum Control},
	author = {Huang, Longxiang and Luneau, Jacquelin and Schirk, Johannes and Wallner, Florian and Schneider, Christian and Filipp, Stefan and Liegener, Klaus and Rabl, Peter},
	year = {2025},
    publisher = {Zenodo},
	doi = {10.5281/zenodo.15574484},
    url = {https://doi.org/10.5281/zenodo.15574484},
}

@misc{xiao2024,
      title={A diagrammatic method to compute the effective Hamiltonian of driven nonlinear oscillators}, 
      author={Xu Xiao and Jayameenakshi Venkatraman and Rodrigo G. Cortiñas and Shoumik Chowdhury and Michel H. Devoret},
      year={2024},
      eprint={2304.13656},
      archivePrefix={arXiv},
      primaryClass={quant-ph},
      url={https://arxiv.org/abs/2304.13656}, 
}

@misc{cornell_2025,
      title={All-mechanical coherence protection and fast control of a spin qubit}, 
      author={Eliza Cornell and Zhujing Xu and Zhaoyou Wang and Hana K. Warner and Eliana Mann and Michael Haas and Smarak Maity and Graham Joe and Liang Jiang and Peter Rabl and Benjamin Pingault and Marko Lončar},
      year={2025},
      eprint={2508.13356},
      archivePrefix={arXiv},
      primaryClass={quant-ph},
      url={https://arxiv.org/abs/2508.13356}, 
}

@misc{glaser_sensitivity-adapted_2024,
	title = {Sensitivity-{adapted} {closed}-{loop} {optimization} for {high}-{fidelity} {Controlled}-{Z} {gates} in {superconducting} {qubits}},
	url = {http://arxiv.org/abs/2412.17454},
	doi = {10.48550/arXiv.2412.17454},
	urldate = {2025-08-13},
	publisher = {arXiv},
	author = {Glaser, Niklas J. and Roy, Federico A. and Tsitsilin, Ivan and Koch, Leon and Bruckmoser, Niklas and Schirk, Johannes and Romeiro, João H. and Huber, Gerhard B. P. and Wallner, Florian and Singh, Malay and Krylov, Gleb and Marx, Achim and Södergren, Lasse and Schneider, Christian M. F. and Werninghaus, Max and Filipp, Stefan},
	month = dec,
	year = {2024},
	note = {arXiv:2412.17454 [quant-ph]},
}

@article{abella_optical_1962,
  title = {Optical Double-Photon Absorption in Cesium Vapor},
  author = {Abella, I. D.},
  journal = {Phys. Rev. Lett.},
  volume = {9},
  issue = {11},
  pages = {453--455},
  numpages = {0},
  year = {1962},
  month = {Dec},
  publisher = {American Physical Society},
  doi = {10.1103/PhysRevLett.9.453},
  url = {https://link.aps.org/doi/10.1103/PhysRevLett.9.453}
}

@article{agostini_multiphoton_1970,
	title = {Multiphoton ionization of rare gases at 1.06 $\mu$ and 0.53 $\mu$},
	volume = {31},
	copyright = {https://www.elsevier.com/tdm/userlicense/1.0/},
	issn = {03759601},
	url = {https://linkinghub.elsevier.com/retrieve/pii/0375960170909874},
	doi = {10.1016/0375-9601(70)90987-4},
	number = {7},
	urldate = {2025-08-13},
	journal = {Phys. Lett. A},
	author = {Agostini, P. and Barjot, G. and Mainfray, G. and Manus, C. and Thebault, J.},
	year = {1970},
	pages = {367--368},
}

@article{bebb_multiphoton_1966,
	title = {Multiphoton {ionization} of {hydrogen} and {rare}-{gas} {atoms}},
	volume = {143},
	copyright = {http://link.aps.org/licenses/aps-default-license},
	issn = {0031-899X},
	url = {https://link.aps.org/doi/10.1103/PhysRev.143.1},
	doi = {10.1103/PhysRev.143.1},
	number = {1},
	urldate = {2025-08-13},
	journal = {Phys. Rev.},
	author = {Bebb, H. Barry and Gold, Albert},
	month = mar,
	year = {1966},
	pages = {1--24},
}

@article{bonch-bruevich_multiphoton_1965,
	title = {{MULTIPHOTON} {PROCESSES}},
	volume = {8},
	issn = {0038-5670},
	url = {https://iopscience.iop.org/article/10.1070/PU1965v008n01ABEH003060},
	doi = {10.1070/PU1965v008n01ABEH003060},
	number = {1},
	urldate = {2025-08-13},
	journal = {Sov. Phys. Uspekhi},
	author = {Bonch-Bruevich, A M and Khodovoĭ, V A},
	month = jan,
	year = {1965},
	pages = {1--38},
}

@article{bravyi_schriefferwolff_2011,
	title = {Schrieffer–{Wolff} transformation for quantum many-body systems},
	volume = {326},
	copyright = {https://www.elsevier.com/tdm/userlicense/1.0/},
	issn = {00034916},
	url = {https://linkinghub.elsevier.com/retrieve/pii/S0003491611001059},
	doi = {10.1016/j.aop.2011.06.004},
	number = {10},
	urldate = {2025-08-13},
	journal = {Ann. Phys. (NY)},
	author = {Bravyi, Sergey and DiVincenzo, David P. and Loss, Daniel},
	month = oct,
	year = {2011},
	pages = {2793--2826},
}

@article{bukov_universal_2015,
	title = {Universal high-frequency behavior of periodically driven systems: From dynamical stabilization to {Floquet} engineering},
	volume = {64},
	issn = {0001-8732, 1460-6976},
	shorttitle = {Universal high-frequency behavior of periodically driven systems},
	url = {http://www.tandfonline.com/doi/full/10.1080/00018732.2015.1055918},
	doi = {10.1080/00018732.2015.1055918},
	number = {2},
	urldate = {2025-08-13},
	journal = {Adv. Phys.},
	author = {Bukov, Marin and D'Alessio, Luca and Polkovnikov, Anatoli},
	month = mar,
	year = {2015},
	pages = {139--226},
}

@article{chakmakjian_observation_1988,
	title = {Observation of resonances at subharmonics of the {Rabi} frequency in the saturated absorption of a 100\% amplitude-modulated laser beam},
	volume = {5},
	copyright = {https://doi.org/10.1364/OA\_License\_v1\#VOR},
	issn = {0740-3224, 1520-8540},
	url = {https://opg.optica.org/abstract.cfm?URI=josab-5-10-2015},
	doi = {10.1364/JOSAB.5.002015},
	number = {10},
	urldate = {2025-08-13},
	journal = {J. Opt. Soc. Am B},
	author = {Chakmakjian, Stephen and Koch, Karl and Stroud, C. R.},
	month = oct,
	year = {1988},
	pages = {2015},
}

@article{didier_ac_2019,
	title = {ac {flux} {sweet} {spots} in {parametrically} {modulated} {superconducting} {qubits}},
	volume = {12},
	issn = {2331-7019},
	url = {https://link.aps.org/doi/10.1103/PhysRevApplied.12.054015},
	doi = {10.1103/PhysRevApplied.12.054015},
	number = {5},
	urldate = {2025-08-13},
	journal = {Phys. Rev. Appl.},
	author = {Didier, Nicolas and Sete, Eyob A. and Combes, Joshua and Da Silva, Marcus P.},
	month = nov,
	year = {2019},
	pages = {054015},
}

@article{ding_high-fidelity_2023,
	title = {High-{fidelity}, {frequency}-{flexible} {two}-{qubit} {fluxonium} {gates} with a {transmon} {coupler}},
	volume = {13},
	issn = {2160-3308},
	url = {https://link.aps.org/doi/10.1103/PhysRevX.13.031035},
	doi = {10.1103/PhysRevX.13.031035},
	number = {3},
	urldate = {2025-08-13},
	journal = {Phys. Rev. X},
	author = {Ding, Leon and Hays, Max and Sung, Youngkyu and Kannan, Bharath and An, Junyoung and Di Paolo, Agustin and Karamlou, Amir H. and Hazard, Thomas M. and Azar, Kate and Kim, David K. and Niedzielski, Bethany M. and Melville, Alexander and Schwartz, Mollie E. and Yoder, Jonilyn L. and Orlando, Terry P. and Gustavsson, Simon and Grover, Jeffrey A. and Serniak, Kyle and Oliver, William D.},
	month = sep,
	year = {2023},
	pages = {031035},
}

@article{eckardt_colloquium_2017,
	title = {Colloquium: {Atomic} quantum gases in periodically driven optical lattices},
	volume = {89},
	copyright = {http://link.aps.org/licenses/aps-default-license},
	issn = {0034-6861, 1539-0756},
	shorttitle = {Colloquium},
	url = {http://link.aps.org/doi/10.1103/RevModPhys.89.011004},
	doi = {10.1103/RevModPhys.89.011004},
	number = {1},
	urldate = {2025-08-13},
	journal = {Rev. Mod. Phys.},
	author = {Eckardt, André},
	month = mar,
	year = {2017},
	pages = {011004},
}

@article{eckardt_high-frequency_2015,
	title = {High-frequency approximation for periodically driven quantum systems from a {Floquet}-space perspective},
	volume = {17},
	issn = {1367-2630},
	url = {https://iopscience.iop.org/article/10.1088/1367-2630/17/9/093039},
	doi = {10.1088/1367-2630/17/9/093039},
	number = {9},
	urldate = {2025-08-13},
	journal = {New J. Phys.},
	author = {Eckardt, André and Anisimovas, Egidijus},
	month = sep,
	year = {2015},
	pages = {093039},
}

@article{goldman_periodically_2014,
	title = {Periodically {driven} {quantum} {systems}: {Effective} {Hamiltonians} and {engineered} {gauge} {fields}},
	volume = {4},
	copyright = {http://creativecommons.org/licenses/by/3.0/},
	issn = {2160-3308},
	shorttitle = {Periodically {Driven} {Quantum} {Systems}},
	url = {https://link.aps.org/doi/10.1103/PhysRevX.4.031027},
	doi = {10.1103/PhysRevX.4.031027},
	number = {3},
	urldate = {2025-08-13},
	journal = {Phys. Rev. X},
	author = {Goldman, N. and Dalibard, J.},
	month = aug,
	year = {2014},
	pages = {031027},
}

@article{goldman_periodically_2015,
	title = {Periodically driven quantum matter: {The} case of resonant modulations},
	volume = {91},
	copyright = {http://link.aps.org/licenses/aps-default-license},
	issn = {1050-2947, 1094-1622},
	shorttitle = {Periodically driven quantum matter},
	url = {https://link.aps.org/doi/10.1103/PhysRevA.91.033632},
	doi = {10.1103/PhysRevA.91.033632},
	number = {3},
	urldate = {2025-08-13},
	journal = {Phys. Rev. A},
	author = {Goldman, N. and Dalibard, J. and Aidelsburger, M. and Cooper, N. R.},
	month = mar,
	year = {2015},
	pages = {033632},
}

@article{gontier_multiphoton_1971,
	title = {Multiphoton {processes} in a {hydrogen} {atom}},
	volume = {4},
	copyright = {http://link.aps.org/licenses/aps-default-license},
	issn = {0556-2791},
	url = {https://link.aps.org/doi/10.1103/PhysRevA.4.1896},
	doi = {10.1103/PhysRevA.4.1896},
	number = {5},
	urldate = {2025-08-13},
	journal = {Phys. Rev. A},
	author = {Gontier, Y. and Trahin, M.},
	month = nov,
	year = {1971},
	pages = {1896--1906},
}

@article{gyenis_experimental_2021,
	title = {Experimental {realization} of a {protected} {superconducting} {circuit} {derived} from the $0–\pi$ {qubit}},
	volume = {2},
	issn = {2691-3399},
	url = {https://link.aps.org/doi/10.1103/PRXQuantum.2.010339},
	doi = {10.1103/PRXQuantum.2.010339},
	number = {1},
	urldate = {2025-08-13},
	journal = {PRX Quantum},
	author = {Gyenis, András and Mundada, Pranav S. and Di Paolo, Agustin and Hazard, Thomas M. and You, Xinyuan and Schuster, David I. and Koch, Jens and Blais, Alexandre and Houck, Andrew A.},
	month = mar,
	year = {2021},
	pages = {010339},
}

@article{gyorgy_electrically_2022,
	title = {Electrically driven spin resonance with bichromatic driving},
	volume = {106},
	issn = {2469-9950, 2469-9969},
	url = {https://link.aps.org/doi/10.1103/PhysRevB.106.155412},
	doi = {10.1103/PhysRevB.106.155412},
	number = {15},
	urldate = {2025-08-13},
	journal = {Phys. Rev. B},
	author = {György, Zoltán and Pályi, András and Széchenyi, Gábor},
	month = oct,
	year = {2022},
	pages = {155412},
}

@article{hertel_gate-tunable_2022,
	title = {Gate-{tunable} {transmon} {using} {selective}-{area}-{grown} {superconductor}-{semiconductor} {hybrid} {structures} on {silicon}},
	volume = {18},
	issn = {2331-7019},
	url = {https://link.aps.org/doi/10.1103/PhysRevApplied.18.034042},
	doi = {10.1103/PhysRevApplied.18.034042},
	 
	number = {3},
	urldate = {2025-08-13},
	journal = {Phys. Rev. Appl.},
	author = {Hertel, Albert and Eichinger, Michaela and Andersen, Laurits O. and Van Zanten, David M.T. and Kallatt, Sangeeth and Scarlino, Pasquale and Kringhøj, Anders and Chavez-Garcia, José M. and Gardner, Geoffrey C. and Gronin, Sergei and Manfra, Michael J. and Gyenis, András and Kjaergaard, Morten and Marcus, Charles M. and Petersson, Karl D.},
	month = sep,
	year = {2022},
	pages = {034042},
}

@article{john_bichromatic_2024,
	title = {Bichromatic {Rabi} {control} of {semiconductor} {qubits}},
	volume = {132},
	issn = {0031-9007, 1079-7114},
	url = {https://link.aps.org/doi/10.1103/PhysRevLett.132.067001},
	doi = {10.1103/PhysRevLett.132.067001},
	number = {6},
	urldate = {2025-08-13},
	journal = {Phys. Rev. Lett.},
	author = {John, Valentin and Borsoi, Francesco and György, Zoltán and Wang, Chien-An and Széchenyi, Gábor and Van Riggelen-Doelman, Floor and Lawrie, William I. L. and Hendrickx, Nico W. and Sammak, Amir and Scappucci, Giordano and Pályi, András and Veldhorst, Menno},
	month = feb,
	year = {2024},
	pages = {067001},
}

@article{koch_subharmonic_1989,
	title = {Subharmonic instabilities in resonant interactions with bichromatic fields},
	volume = {6},
	copyright = {https://doi.org/10.1364/OA\_License\_v1\#VOR},
	issn = {0740-3224, 1520-8540},
	url = {https://opg.optica.org/abstract.cfm?URI=josab-6-1-58},
	doi = {10.1364/JOSAB.6.000058},
	number = {1},
	urldate = {2025-08-13},
	journal = {J. Opt. Soc. Am. B},
	author = {Koch, Karl and Hillman, Lloyd W. and Oliver, Brian J. and Chakmakjian, Stephen H. and Stroud, C. R.},
	month = jan,
	year = {1989},
	pages = {58},
}

@article{lambropoulos_topics_1976,
	title = {Topics on Multiphoton Processes in Atoms},
	volume = {12},
	isbn = {978-0-12-003812-1},
	copyright = {},
	url = {https://linkinghub.elsevier.com/retrieve/pii/S0065219908600433},
	doi = {10.1016/S0065-2199(08)60043-3},
	urldate = {2025-08-13},
	journal = {Adv. At. Mol. Phys},
	publisher = {Elsevier},
	author = {Lambropoulos, P.},
	year = {1976},
	pages = {87--164},
}

@article{lambropoulos_two-electron_1998,
	title = {Two-electron atoms in strong fields},
	volume = {305},
	copyright = {https://www.elsevier.com/tdm/userlicense/1.0/},
	issn = {03701573},
	url = {https://linkinghub.elsevier.com/retrieve/pii/S0370157398000271},
	doi = {10.1016/S0370-1573(98)00027-1},
	number = {5},
	urldate = {2025-08-13},
	journal = {Phys. Rep.},
	author = {Lambropoulos, P. and Maragakis, P. and Zhang, Jian},
	month = nov,
	year = {1998},
	pages = {203--293},
}

@article{magnus_exponential_1954,
	title = {On the exponential solution of differential equations for a linear operator},
	volume = {7},
	issn = {0010-3640, 1097-0312},
	url = {https://onlinelibrary.wiley.com/doi/10.1002/cpa.3160070404},
	doi = {10.1002/cpa.3160070404},
	number = {4},
	urldate = {2025-08-13},
	journal = {Commun. Pure Appl. Math},
	author = {Magnus, Wilhelm},
	month = nov,
	year = {1954},
	pages = {649--673},
}

@article{mikami_brillouin-wigner_2016,
	title = {Brillouin-{Wigner} theory for high-frequency expansion in periodically driven systems: {Application} to {Floquet} topological insulators},
	volume = {93},
	copyright = {http://link.aps.org/licenses/aps-default-license},
	issn = {2469-9950, 2469-9969},
	shorttitle = {Brillouin-{Wigner} theory for high-frequency expansion in periodically driven systems},
	url = {https://link.aps.org/doi/10.1103/PhysRevB.93.144307},
	doi = {10.1103/PhysRevB.93.144307},
	number = {14},
	urldate = {2025-08-13},
	journal = {Phys. Rev. B},
	author = {Mikami, Takahiro and Kitamura, Sota and Yasuda, Kenji and Tsuji, Naoto and Oka, Takashi and Aoki, Hideo},
	month = apr,
	year = {2016},
	pages = {144307},
}

@article{rodriguez-vega_low-frequency_2021,
	title = {Low-frequency and {Moiré}–{Floquet} engineering: {A} review},
	volume = {435},
	issn = {00034916},
	shorttitle = {Low-frequency and {Moiré}–{Floquet} engineering},
	url = {https://linkinghub.elsevier.com/retrieve/pii/S0003491621000403},
	doi = {10.1016/j.aop.2021.168434},
	urldate = {2025-08-13},
	journal = {Ann. Phys. (NY)},
	author = {Rodriguez-Vega, Martin and Vogl, Michael and Fiete, Gregory A.},
	month = dec,
	year = {2021},
	pages = {168434},
}

@article{romhanyi_subharmonic_2015,
	title = {Subharmonic transitions and {Bloch}-{Siegert} shift in electrically driven spin resonance},
	volume = {92},
	copyright = {http://link.aps.org/licenses/aps-default-license},
	issn = {1098-0121, 1550-235X},
	url = {https://link.aps.org/doi/10.1103/PhysRevB.92.054422},
	doi = {10.1103/PhysRevB.92.054422},
	number = {5},
	urldate = {2025-08-13},
	journal = {Phys. Rev. B},
	author = {Romhányi, Judit and Burkard, Guido and Pályi, András},
	month = aug,
	year = {2015},
	pages = {054422},
}

@article{rosenfeld_high-fidelity_2024,
	title = {High-{fidelity} {two}-{qubit} {gates} between {fluxonium} {qubits} with a {resonator} {coupler}},
	volume = {5},
	issn = {2691-3399},
	url = {https://link.aps.org/doi/10.1103/PRXQuantum.5.040317},
	doi = {10.1103/PRXQuantum.5.040317},
	 
	number = {4},
	urldate = {2025-08-13},
	journal = {PRX Quantum},
	author = {Rosenfeld, Emma L. and Hann, Connor T. and Schuster, David I. and Matheny, Matthew H. and Clerk, Aashish A.},
	month = nov,
	year = {2024},
	pages = {040317},
}

@article{rudolph_shift_1998,
	title = {Shift of the subharmonic resonances and suppression of fluorescence in a two-level atom driven by a bichromatic field},
	volume = {15},
	copyright = {https://doi.org/10.1364/OA\_License\_v1\#VOR},
	issn = {0740-3224, 1520-8540},
	url = {https://opg.optica.org/abstract.cfm?URI=josab-15-8-2325},
	doi = {10.1364/JOSAB.15.002325},
	number = {8},
	urldate = {2025-08-13},
	journal = {J. Opt. Soc. Am. B},
	author = {Rudolph, T. and Freedhoff, H. S. and Ficek, Z.},
	month = aug,
	year = {1998},
	pages = {2325},
}

@article{rudolph_multiphoton_1998,
	title = {Multiphoton ac {Stark} effect in a bichromatically driven two-level atom},
	volume = {58},
	copyright = {http://link.aps.org/licenses/aps-default-license},
	issn = {1050-2947, 1094-1622},
	url = {https://link.aps.org/doi/10.1103/PhysRevA.58.1296},
	doi = {10.1103/PhysRevA.58.1296}, 
	number = {2},
	urldate = {2025-08-13},
	journal = {Phys. Rev. A},
	author = {Rudolph, T. G. and Freedhoff, H. S. and Ficek, Z.},
	month = aug,
	year = {1998},
	pages = {1296--1309},
}

@article{rudolph_multiphoton_1998-1,
	title = {The multiphoton {AC} {Stark} effect},
	volume = {147},
	copyright = {https://www.elsevier.com/tdm/userlicense/1.0/},
	issn = {00304018},
	url = {https://linkinghub.elsevier.com/retrieve/pii/S0030401897006548},
	doi = {10.1016/S0030-4018(97)00654-8},
	number = {1-3},
	urldate = {2025-08-13},
	journal = {Opt. Commun.},
	author = {Rudolph, T.G and Ficek, Z and Freedhoff, H.S},
	month = feb,
	year = {1998},
	pages = {78--82},
}

@article{ruyten_subharmonic_1989,
	title = {Subharmonic resonances in terms of multiphoton processes and generalized {Bloch}–{Siegert} shifts},
	volume = {6},
	copyright = {https://doi.org/10.1364/OA\_License\_v1\#VOR},
	issn = {0740-3224, 1520-8540},
	url = {https://opg.optica.org/abstract.cfm?URI=josab-6-10-1796},
	doi = {10.1364/JOSAB.6.001796},
	number = {10},
	urldate = {2025-08-13},
	journal = {J. Opt. Soc. Am. B},
	author = {Ruyten, Wilhelmus M.},
	month = oct,
	year = {1989},
	pages = {1796},
}

@article{sah_decay-protected_2024,
	title = {Decay-protected superconducting qubit with fast control enabled by integrated on-chip filters},
	volume = {7},
	issn = {2399-3650},
	url = {https://www.nature.com/articles/s42005-024-01733-3},
	doi = {10.1038/s42005-024-01733-3},
	 
	number = {1},
	urldate = {2025-08-13},
	journal = {Commun. Phys.},
	author = {Sah, Aashish and Kundu, Suman and Suominen, Heikki and Chen, Qiming and Möttönen, Mikko},
	month = jul,
	year = {2024},
	pages = {227},
}

@article{sambe_steady_1973,
	title = {Steady {states} and {quasienergies} of a {quantum}-{mechanical} {system} in an {oscillating} {field}},
	volume = {7},
	copyright = {http://link.aps.org/licenses/aps-default-license},
	issn = {0556-2791},
	url = {https://link.aps.org/doi/10.1103/PhysRevA.7.2203},
	doi = {10.1103/PhysRevA.7.2203},
	number = {6},
	urldate = {2025-08-13},
	journal = {Phys. Rev. A},
	author = {Sambe, Hideo},
	month = jun,
	year = {1973},
	pages = {2203--2213},
}

@article{sarkar_subharmonics_2021,
	title = {Subharmonics and superharmonics of the weak field in a driven two-level quantum system: {Vibrational} resonance enhancement},
	volume = {104},
	issn = {2470-0045, 2470-0053},
	shorttitle = {Subharmonics and superharmonics of the weak field in a driven two-level quantum system},
	url = {https://link.aps.org/doi/10.1103/PhysRevE.104.014202},
	doi = {10.1103/PhysRevE.104.014202},
	number = {1},
	urldate = {2025-08-13},
	journal = {Phys. Rev. E},
	author = {Sarkar, Prasun and Paul, Shibashis and Ray, Deb Shankar},
	month = jul,
	year = {2021},
	pages = {014202},
}

@article{schirk_subharmonic_2025,
	title = {Subharmonic {control} of a {fluxonium} {qubit} via a {Purcell}-{protected} {flux} {line}},
	volume = {6},
	issn = {2691-3399},
	url = {https://link.aps.org/doi/10.1103/yx15-jyl7},
	doi = {10.1103/yx15-jyl7},
	 
	number = {3},
	urldate = {2025-08-13},
	journal = {PRX Quantum},
	author = {Schirk, J. and Wallner, F. and Huang, L. and Tsitsilin, I. and Bruckmoser, N. and Koch, L. and Bunch, D. and Glaser, N.J. and Huber, G.B.P. and Knudsen, M. and Krylov, G. and Marx, A. and Pfeiffer, F. and Richard, L. and Roy, F.A. and Romeiro, J.H. and Singh, M. and Södergren, L. and Dionis, E. and Sugny, D. and Werninghaus, M. and Liegener, K. and Schneider, C.M.F. and Filipp, S.},
	month = jul,
	year = {2025},
	pages = {030315},
}

@article{schrieffer_relation_1966,
	title = {Relation between the {Anderson} and {Kondo} {Hamiltonians}},
	volume = {149},
	copyright = {http://link.aps.org/licenses/aps-default-license},
	issn = {0031-899X},
	url = {https://link.aps.org/doi/10.1103/PhysRev.149.491},
	doi = {10.1103/PhysRev.149.491},
	number = {2},
	urldate = {2025-08-13},
	journal = {Phys. Rev.},
	author = {Schrieffer, J. R. and Wolff, P. A.},
	month = sep,
	year = {1966},
	pages = {491--492},
}

@article{schweizer_floquet_2019,
	title = {Floquet approach to $\mathbb{Z}_2$ lattice gauge theories with ultracold atoms in optical lattices},
	volume = {15},
	issn = {1745-2473, 1745-2481},
	url = {https://www.nature.com/articles/s41567-019-0649-7},
	doi = {10.1038/s41567-019-0649-7},
	number = {11},
	urldate = {2025-08-13},
	journal = {Nat. Phys.},
	author = {Schweizer, Christian and Grusdt, Fabian and Berngruber, Moritz and Barbiero, Luca and Demler, Eugene and Goldman, Nathan and Bloch, Immanuel and Aidelsburger, Monika},
	month = nov,
	year = {2019},
	pages = {1168--1173},
}

@article{shirley_solution_1965,
	title = {Solution of the {Schrödinger} {Equation} with a {Hamiltonian} {Periodic} in {Time}},
	volume = {138},
	copyright = {http://link.aps.org/licenses/aps-default-license},
	issn = {0031-899X},
	url = {https://link.aps.org/doi/10.1103/PhysRev.138.B979},
	doi = {10.1103/PhysRev.138.B979},
	number = {4B},
	urldate = {2025-08-13},
	journal = {Phys. Rev.},
	author = {Shirley, Jon H.},
	month = may,
	year = {1965},
	pages = {B979--B987},
}

@article{soliverez_effective_1969,
	title = {An effective {Hamiltonian} and time-independent perturbation theory},
	volume = {2},
	issn = {00223719},
	url = {https://iopscience.iop.org/article/10.1088/0022-3719/2/12/301},
	doi = {10.1088/0022-3719/2/12/301},
	number = {12},
	urldate = {2025-08-13},
	journal = {J. Phys. C: Solid State Phys.},
	author = {Soliverez, C E},
	month = dec,
	year = {1969},
	pages = {2161--2174},
}

@article{suzuki_degenerate_1983,
	title = {Degenerate {perturbation} {theory} in {quantum} {mechanics}},
	volume = {70},
	issn = {0033-068X, 1347-4081},
	url = {https://academic.oup.com/ptp/article-lookup/doi/10.1143/PTP.70.439},
	doi = {10.1143/PTP.70.439},
	number = {2},
	urldate = {2025-08-13},
	journal = {Prog. Theor. Phys.},
	author = {Suzuki, K. and Okamoto, R.},
	month = aug,
	year = {1983},
	pages = {439--451},
}

@article{szechenyi_maximal_2014,
	title = {Maximal {Rabi} frequency of an electrically driven spin in a disordered magnetic field},
	volume = {89},
	copyright = {http://link.aps.org/licenses/aps-default-license},
	issn = {1098-0121, 1550-235X},
	url = {https://link.aps.org/doi/10.1103/PhysRevB.89.115409},
	doi = {10.1103/PhysRevB.89.115409},
	number = {11},
	urldate = {2025-08-13},
	journal = {Phys. Rev. B},
	author = {Széchenyi, Gábor and Pályi, András},
	month = mar,
	year = {2014},
	pages = {115409},
}

@article{villafane_three-photon_2023,
	title = {Three-{photon} {excitation} of {InGaN} {quantum} {dots}},
	volume = {130},
	issn = {0031-9007, 1079-7114},
	url = {https://link.aps.org/doi/10.1103/PhysRevLett.130.083602},
	doi = {10.1103/PhysRevLett.130.083602},
	number = {8},
	urldate = {2025-08-13},
	journal = {Phys. Rev. Lett.},
	author = {Villafañe, Viviana and Scaparra, Bianca and Rieger, Manuel and Appel, Stefan and Trivedi, Rahul and Zhu, Tongtong and Jarman, John and Oliver, Rachel A. and Taylor, Robert A. and Finley, Jonathan J. and Müller, Kai},
	month = feb,
	year = {2023},
	pages = {083602},
}

@article{werninghaus_leakage_2021,
	title = {Leakage reduction in fast superconducting qubit gates via optimal control},
	volume = {7},
	issn = {2056-6387},
	url = {https://www.nature.com/articles/s41534-020-00346-2},
	doi = {10.1038/s41534-020-00346-2},
	number = {1},
	urldate = {2025-08-13},
	journal = {npj Quantum Inf.},
	author = {Werninghaus, M. and Egger, D. J. and Roy, F. and Machnes, S. and Wilhelm, F. K. and Filipp, S.},
	month = jan,
	year = {2021},
	pages = {14},
}

@misc{heunisch_scalable_2025,
	title = {Scalable {fluxonium}-{transmon} {architecture} for {error} {corrected} {quantum} {processors}},
	url = {http://arxiv.org/abs/2508.09267},
	doi = {10.48550/arXiv.2508.09267},
	urldate = {2025-08-14},
	publisher = {arXiv},
	author = {Heunisch, Lukas and Huang, Longxiang and Tasler, Stephan and Schirk, Johannes and Wallner, Florian and Feulner, Verena and Sarma, Bijita and Liegener, Klaus and Schneider, Christian M. F. and Filipp, Stefan and Hartmann, Michael J.},
	month = aug,
	year = {2025},
	note = {arXiv:2508.09267 [quant-ph]},
}

@article{xia_fast_2025,
  title = {Fast superconducting qubit control with subharmonic drives},
  author = {Xia, Mingkang and Zhou, Chao and Liu, Chenxu and Patel, Param and Cao, Xi and Lu, Pinlei and Mesits, Boris and Mucci, Maria and Gorski, David and Pekker, David and Hatridge, Michael},
  year = 2025,
  journal = {Nature Communications},
  volume = {17},
  number = {1},
  pages = {1024},
  issn = {2041-1723},
  doi = {10.1038/s41467-025-67766-6}
}

@article{young_third-harmonic_1971,
	title = {Third-{harmonic} {generation} in {phase}-{matched} {Rb} {vapor}},
	volume = {27},
	copyright = {http://link.aps.org/licenses/aps-default-license},
	issn = {0031-9007},
	url = {https://link.aps.org/doi/10.1103/PhysRevLett.27.1551},
	doi = {10.1103/PhysRevLett.27.1551},
	number = {23},
	urldate = {2025-08-14},
	journal = {Phys. Rev. Lett.},
	author = {Young, J. F. and Bjorklund, G. C. and Kung, A. H. and Miles, R. B. and Harris, S. E.},
	month = dec,
	year = {1971},
	pages = {1551--1553},
}

@article{you_circuit_2019,
	title = {Circuit quantization in the presence of time-dependent external flux},
	volume = {99},
	issn = {2469-9950, 2469-9969},
	url = {https://link.aps.org/doi/10.1103/PhysRevB.99.174512},
	doi = {10.1103/PhysRevB.99.174512},
	number = {17},
	urldate = {2025-08-14},
	journal = {Phys. Rev. B},
	author = {You, Xinyuan and Sauls, J. A. and Koch, Jens},
	month = may,
	year = {2019},
	pages = {174512},
}

@article{xu_perturbative_2025,
	title = {Perturbative framework for engineering arbitrary {Floquet} {Hamiltonian}},
	volume = {88},
	issn = {0034-4885, 1361-6633},
	url = {https://iopscience.iop.org/article/10.1088/1361-6633/adb072},
	doi = {10.1088/1361-6633/adb072},
	number = {3},
	urldate = {2025-08-14},
	journal = {Rep. Prog. Phys.},
	author = {Xu, Yingdan and Guo, Lingzhen},
	month = mar,
	year = {2025},
	pages = {037602},
}

@article{oliver_mach-zehnder_2005,
	title = {Mach-{Zehnder} {Iinterferometry} in a {strongly} {driven} {superconducting} {qubit}},
	volume = {310},
	issn = {0036-8075, 1095-9203},
	url = {https://www.science.org/doi/10.1126/science.1119678},
	doi = {10.1126/science.1119678},
	number = {5754},
	urldate = {2025-08-20},
	journal = {Science},
	author = {Oliver, William D. and Yu, Yang and Lee, Janice C. and Berggren, Karl K. and Levitov, Leonid S. and Orlando, Terry P.},
	month = dec,
	year = {2005},
	pages = {1653--1657},
}

@misc{cornell_all-mechanical_2025,
	title = {All-mechanical coherence protection and fast control of a spin qubit},
	url = {http://arxiv.org/abs/2508.13356},
	doi = {10.48550/arXiv.2508.13356},
	urldate = {2025-08-22},
	publisher = {arXiv},
	author = {Cornell, Eliza and Xu, Zhujing and Wang, Zhaoyou and Warner, Hana K. and Mann, Eliana and Haas, Michael and Maity, Smarak and Joe, Graham and Jiang, Liang and Rabl, Peter and Pingault, Benjamin and Lončar, Marko},
	month = aug,
	year = {2025},
	note = {arXiv:2508.13356 [quant-ph]},
}

@article{deppe_two-photon_2008,
	title = {Two-photon probe of the {Jaynes}–{Cummings} model and controlled symmetry breaking in circuit {QED}},
	volume = {4},
	issn = {1745-2473, 1745-2481},
	url = {https://www.nature.com/articles/nphys1016},
	doi = {10.1038/nphys1016},
	number = {9},
	urldate = {2025-08-22},
	journal = {Nat. Phys.},
	author = {Deppe, Frank and Mariantoni, Matteo and Menzel, E. P. and Marx, A. and Saito, S. and Kakuyanagi, K. and Tanaka, H. and Meno, T. and Semba, K. and Takayanagi, H. and Solano, E. and Gross, R.},
	month = sep,
	year = {2008},
	pages = {686--691},
}

@article{gyorgy_limitations_2025,
	title = {Limitations of the g -tensor formalism of semiconductor spin qubits},
	volume = {112},
	issn = {2469-9950, 2469-9969},
	url = {https://link.aps.org/doi/10.1103/gq3h-76mx},
	doi = {10.1103/gq3h-76mx},
	number = {4},
	urldate = {2025-08-26},
	journal = {Phys. Rev. B},
	author = {György, Zoltán and Pályi, András and Széchenyi, Gábor},
	month = jul,
	year = {2025},
	pages = {045428},
}

@article{ho1983SemiclassicalManymodeFloquet,
  title = {Semiclassical many-mode Floquet theory},
  author = {Ho, Tak-San and Chu, Shih-I and Tietz, James V.},
  year = {1983},
  journal = {Chem. Phys. Lett.},
  volume = {96},
  number = {4},
  pages = {464--471},
  issn = {00092614},
  doi = {10.1016/0009-2614(83)80732-5}
}

@article{chu2004FloquetTheoremGeneralizeda,
  title = {Beyond the {{Floquet}} Theorem: Generalized {{Floquet}} Formalisms and Quasienergy Methods for Atomic and Molecular Multiphoton Processes in Intense Laser Fields},
  shorttitle = {Beyond the {{Floquet}} Theorem},
  author = {Chu, Shih-I and Telnov, Dmitry A.},
  year = {2004},
  journal = {Phys. Rep.},
  volume = {390},
  number = {1-2},
  pages = {1--131},
  issn = {03701573},
  doi = {10.1016/j.physrep.2003.10.001},
  copyright = {https://www.elsevier.com/tdm/userlicense/1.0/}
}

@article{munoz2018FilteringMultiphotonEmission,
  title = {Filtering Multiphoton Emission from State-of-the-Art Cavity Quantum Electrodynamics},
  author = {Mu{\~n}oz, C. S{\'a}nchez and Laussy, Fabrice P. and Valle, Elena Del and Tejedor, Carlos and {Gonz{\'a}lez-Tudela}, Alejandro},
  year = {2018},
  journal = {Optica},
  volume = {5},
  number = {1},
  pages = {14},
  issn = {2334-2536},
  doi = {10.1364/OPTICA.5.000014}
}

@article{munoz2014EmittersNphotonBundles,
  title = {Emitters of {{$N$-photon}} Bundles},
  author = {Mu{\~n}oz, C. S{\'a}nchez and Del Valle, E. and Tudela, A. Gonz{\'a}lez and M{\"u}ller, K. and Lichtmannecker, S. and Kaniber, M. and Tejedor, C. and Finley, J. J. and Laussy, F. P.},
  year = {2014},
  journal = {Nat. Photonics},
  volume = {8},
  number = {7},
  pages = {550--555},
  issn = {1749-4885, 1749-4893},
  doi = {10.1038/nphoton.2014.114}
}

@article{shevchenko2012MultiphotonTransitionsJosephsonjunction,
  title = {Multiphoton transitions in {{Josephson-junction}} Qubits ({{review article}})},
  author = {Shevchenko, S. N. and Omelyanchouk, A. N. and Il'ichev, E.},
  year = {2012},
  journal = {Low Temperature Physics},
  volume = {38},
  number = {4},
  pages = {283--300},
  issn = {1063-777X, 1090-6517},
  doi = {10.1063/1.3701717}
}

@article{shevchenko2010LandauZenerStuckelbergInterferometry,
  title = {Landau-{{Zener-St{\"u}ckelberg}} Interferometry},
  author = {Shevchenko, Sergey N. and Ashhab, S. and Nori, Franco},
  year = {2010},
  journal = {Physics Reports},
  volume = {492},
  number = {1},
  pages = {1--30},
  issn = {03701573},
  doi = {10.1016/j.physrep.2010.03.002}
}

@article{ivakhnenko2023NonadiabaticLandauZener,
  title = {Nonadiabatic {{Landau}}--{{Zener}}--{{St{\"u}ckelberg}}--{{Majorana}} Transitions, Dynamics, and Interference},
  author = {Ivakhnenko, Oleh V. and Shevchenko, Sergey N. and Nori, Franco},
  year = {2023},
  journal = {Physics Reports},
  volume = {995},
  pages = {1--89},
  issn = {03701573},
  doi = {10.1016/j.physrep.2022.10.002}
}

@article{xiong2022ArbitraryControlledphaseGate,
  title = {Arbitrary controlled-Phase gate on fluxonium qubits using differential ac {{Stark}} shifts},
  author = {Xiong, Haonan and Ficheux, Quentin and Somoroff, Aaron and Nguyen, Long B. and Dogan, Ebru and Rosenstock, Dario and Wang, Chen and Nesterov, Konstantin N. and Vavilov, Maxim G. and Manucharyan, Vladimir E.},
  year = 2022,
  journal = {Physical Review Research},
  volume = {4},
  number = {2},
  pages = {023040},
  issn = {2643-1564},
  doi = {10.1103/PhysRevResearch.4.023040}
}

@article{nesterov2021ProposalEntanglingGates,
  title = {Proposal for {{Entangling Gates}} on {{Fluxonium Qubits}} via a {{Two-Photon Transition}}},
  author = {Nesterov, Konstantin N. and Ficheux, Quentin and Manucharyan, Vladimir E. and Vavilov, Maxim G.},
  year = 2021,
  journal = {PRX Quantum},
  volume = {2},
  number = {2},
  pages = {020345},
  issn = {2691-3399},
  doi = {10.1103/PRXQuantum.2.020345}
}

@article{rahav2003Effective,
  title = {Effective {{Hamiltonians}} for Periodically Driven Systems},
  author = {Rahav, Saar and Gilary, Ido and Fishman, Shmuel},
  year = 2003,
  journal = {Physical Review A},
  volume = {68},
  number = {1},
  pages = {013820},
  issn = {1050-2947, 1094-1622},
  doi = {10.1103/PhysRevA.68.013820},
  copyright = {http://link.aps.org/licenses/aps-default-license}
}

@article{casas2001Floquet,
  title = {Floquet Theory: Exponential Perturbative Treatment},
  shorttitle = {Floquet Theory},
  author = {Casas, F and Oteo, J A and Ros, J},
  year = 2001,
  journal = {Journal of Physics A: Mathematical and General},
  volume = {34},
  number = {16},
  pages = {3379--3388},
  issn = {0305-4470, 1361-6447},
  doi = {10.1088/0305-4470/34/16/305}
}

@article{venkatraman2022Static,
  title = {On the Static Effective {{Hamiltonian}} of a Rapidly Driven Nonlinear System},
  author = {Venkatraman, Jayameenakshi and Xiao, Xu and Corti{\~n}as, Rodrigo G. and Eickbusch, Alec and Devoret, Michel H.},
  year = 2022,
  journal = {Physical Review Letters},
  volume = {129},
  number = {10},
  pages = {100601},
  issn = {0031-9007, 1079-7114},
  doi = {10.1103/PhysRevLett.129.100601},
}

@article{dey2025ErrorBoundsFloquetMagnus,
  title = {Error Bounds for the {{Floquet-Magnus}} Expansion and Their Application to the Semiclassical Quantum {{Rabi}} Model},
  author = {Dey, Anirban and Lonigro, Davide and Yuasa, Kazuya and Burgarth, Daniel},
  year = 2025,
  journal = {Physical Review A},
  volume = {112},
  number = {5},
  pages = {053723},
  issn = {2469-9926, 2469-9934},
  doi = {10.1103/6bgj-s987}
}

@article{klein1974DegeneratePerturbationTheory,
  title = {Degenerate Perturbation Theory},
  author = {Klein, D. J.},
  year = 1974,
  journal = {The Journal of Chemical Physics},
  volume = {61},
  number = {3},
  pages = {786--798},
  issn = {0021-9606, 1089-7690},
  doi = {10.1063/1.1682018}
}
\appendix

\section{Degenerate Perturbation Theory}\label{app:DPT}

This appendix outlines the derivation of the effective Hamiltonian from degenerate perturbation theory, following the general analysis of Ref.~\cite{soliverez_effective_1969}.
We apply it to the case of the Floquet-Sambe Hamiltonian
\begin{align}
    \H &= \H_0 + \V,
\end{align}
with a $d$-fold degenerate eigenvalue~$\tilde E_0$ of the unperturbed Hamiltonian~$\H_0$. As a starting point, we use the decomposition of the system into the $P$ and $Q$ subspaces introduced in Sec.~\ref{sec:DPT}, and we restate once more the eigenvalue equations 
\begin{align}\label{eq:EigenvalueEq_App_full} 
    \H\ket{A} &= (\H_0 + \V)\ket{A}=  (\tilde E_0+\delta E_A)\ket{A},\\
    \H_\eff \ket{A^\prime } &= \delta E_A \ket{A^\prime}, \label{eq:EigenvalueEq_App_eff}
\end{align}
which determine $\H_\eff=W^\dag (\H-\tilde E_0) W $ and the transformation $W$ through $|A^\prime\rangle= W^\dag |A\rangle$.



As long as $\V$ is sufficiently weak, the perturbed eigenenergies $\delta E_A$, with $\delta E_A(\lambda\rightarrow 0 )=0$, and the transformation $W$ can be derived in perturbation theory. To do so, we first project Eq.~\eqref{eq:EigenvalueEq_App_full} onto the $P$ and $Q$ subspaces and obtain
\begin{align}
P \V  |A\rangle= & \,  \delta E_A |A_0\rangle,  \label{eq:projectedP}  \\ 
|A\rangle= (P+Q)  |A\rangle = \,& |A_0\rangle -  R( \delta E_A- \V) |A\rangle, \label{eq:projectedPQ}
\end{align}
where $|A_0\rangle=P|A\rangle$ and we have introduced the resolvent operator
\begin{align}\label{eq:resolvent}
    R = Q\frac{1}{\tilde E_0-\H_0} Q
    &= \sum_{p=-\infty}^\infty \sum_{\substack{l\notin D_p }} \frac{\kket{l,p}\bbra{l,p}}{\tilde E_0 +p\omega_d - \tilde E_l}.
\end{align}
Here $D_p=\{\ket{k}\in D ~|~ n_k = p\}$ is the ensemble of $p$-photon resonant states, 
such that the restriction in the sum keeps only non-diverging terms. In a next step, we define the so-called wave operator $L$~\cite{soliverez_effective_1969,suzuki_degenerate_1983} via the relation 
\begin{equation}
|A\rangle= L |A_0\rangle.
\end{equation}
This operator maps the projection $|A_0\rangle$ back into the full Hilbert space and vanishes when acting on the complementary subspace, $L=LP$. By combining Eq.~\eqref{eq:projectedP} and Eq.~\eqref{eq:projectedPQ}, we find that it satisfies~\cite{soliverez_effective_1969}
\begin{equation}\label{eq:app:Ldef}
    L = P + R\V L - RL\V L.
\end{equation}
With the help of the wave operator, we can convert Eq.~\eqref{eq:projectedP} into the eigenvalue equation 
\begin{equation}\label{eq:eigenvalue_nonHermitian} 
P \V L |A_0\rangle = \delta E_A |A_0\rangle,
\end{equation} 
where, however the operator $P \V L$ is not Hermitian. Therefore, in a final step, we define 
\begin{equation}\label{eq:app:Ndef}
    N = L^\dagger L,
\end{equation}
which is a positive definite operator acting on the degenerate subspace only, $N=NP=PN$~\cite{soliverez_effective_1969}. Therefore,  we can define the operators $N^{1/2}$ and $N^{-1/2}$ within this subspace. By multiplying Eq.~\eqref{eq:eigenvalue_nonHermitian} with $N^{1/2}$ from the left and setting $|A^\prime\rangle = N^{1/2}|A_0\rangle$, we obtain the targeted Eq.~\eqref{eq:EigenvalueEq_App_eff} with a Hermitian effective Hamiltonian 
\begin{equation}\label{eq:app:Heffdef}
   \H_\eff= PN^{1/2}\V LN^{-1/2}P
\end{equation}
and the corresponding transformation
\begin{align}
      W = LN^{-1/2}P, \label{eq:app:W_DPT}
\end{align}
between the eigenstates $|A^\prime\rangle$ in $D$ and the eigenstate $|A\rangle$ in the full Sambe space.

The remaining task is to  express the effective Hamiltonian in powers of~$\V$, 
\begin{equation}
    \H_\eff = \sum_{r=0}^\infty \H_{\eff}^{(r)},
\end{equation}
where~$\H_{\eff}^{(r)}$ is of order~$\lambda^r$. A  similar expansion with coefficients $L_r$, $N_r$, $N_r^{\pm1/2}$, and $W_r\equiv W^{(r)}$ is also assumed for the other operators. 
The recurrence relation between the different orders is obtained by inserting these expansions into Eqs.~\eqref{eq:app:Ldef}, \eqref{eq:app:Ndef}, \eqref{eq:app:Heffdef}, and \eqref{eq:app:W_DPT} as well as in the relations $N^{1/2}N^{1/2}=N$ and $N^{-1/2}N^{1/2}=P$.
To zeroth order we obtain $L_0=N_0^{1/2}=N_0^{-1/2} =W_0 =P$ and $\H_{\eff}^{(0)}= 0$. For the next orders, we find that $N_1^{1/2} = 0$ and $L_1 = R\V P$, while all other contributions can be obtained in an iterative manner by solving the recurrence relations
\begin{align}
    L_r &= R\V L_{r-1} - \sum_{k=1}^{r-1}RL_k\V L_{r-k-1} , \\
    N_r &= \sum_{k=0}^r L^\dagger_k L_{r-k} , \\
    N_r^{1/2} &= \frac{1}{2}N_r-\frac{1}{2}\sum_{k=1}^{r-1}N_k^{1/2}N_{r-k}^{1/2}, \\
    N_r^{-1/2} &= -\sum_{k=0}^{r-1}N_k^{-1/2}N_{r-k}^{1/2}, \\
    W_r &= \sum_{k=0}^r L_k N_{r-k}^{-1/2} , \\
    \H_{\eff}^{(r)} &= \sum_{k=0}^{r-1} \sum_{l=0}^{r-1-k} N^{1/2}_k \V L_l N^{-1/2}_{r-k-l-1}.
\end{align}

From these recurrence relations, we see that each operator of order~$r$ can be written as a linear combination of \textit{operator strings} of the form
\begin{equation}\label{eq:app:opp_string}
    R^{m_r}V\dots R^{m_1}V R^{m_0},
\end{equation}
where $m_j\geq 0$ and $R^0\equiv P$.
The effective Hamiltonian acts only in the degenerate subspace, such that the first and last operators are~$P$, $m_0=m_r=0$. The explicit expression for the lowest orders is
\begin{align}
    \H_\eff^{(1)} &= P\V P, \label{eq:Heff1}\\
    \H_\eff^{(2)} &= P\V R\V P, \label{eq:Heff2}\\
    \H_\eff^{(3)} &=  P \V R \V R \V P-\frac{1}{2}\left( P \V P \V R^2\V P  +  P \V R^2\V P \V P \right). \label{eq:Heff3}
\end{align}
In general, the $r$-th order effective Hamiltonian is of the form
\begin{equation}\label{eq:Heff_r}
    \H_\eff^{(r)} = \sum_{\substack{m_1 + \dots + m_{r-1} = r-1}} c_{m_{r-1}, \dots, m_{1}} P\V R^{m_{r-1}}\dots R^{m_1}\V P,
\end{equation}
where $R^m \equiv \left(\frac{1-P}{\tilde E_0 - \H_0}\right)^m$ for $m\geq 1$.
The coefficients $c_{m_1, \dots, m_{r-1}}$ are numbers determined by solving above equations recursively and whose
non-zero terms are summarized in Table~\ref{table:DPTcoeff_order3-4} for $r=3$ and $r=4$, and in Table~\ref{table:DPTcoeff_order5} for $r=5$. At all orders the simplest type of process $P\V R\dots \V P$ corresponding to the exponents $m_1=\dots=m_{r-1}=1$ and has a trivial coefficient $c_{1,\dots,1}=1$.
Note that these coefficients are independent of the Hamiltonian and dimension of the degenerate subspace. 

\corr{
To obtain the expression for the matrix elements of the effective Hamiltonian given in Eq.~\eqref{eq:Heff_matEl_general}, we first apply $\kket{k,n_k}$ on the right of Eq.~\eqref{eq:Heff_r} and expand each $\V\kket{a,q}=\sum_{p,\ket{a'}}\kket{a',p+q}V_{p,a'a}$.
Applying $\bbra{l,n_l}$ on the left leads to the energy constraint in Eq.~\eqref{eq:harmonicsSum}. 
We then use Eq.~\eqref{eq:resolvent} for the resolvent.
In particular, for $m>0$, we have $R^m\kket{a,q+n_k}=0$ if $\kket{a,q+n_k}$ belongs to the degenerate subspace, i.e. if $\tilde E_a=\tilde E_k+q\omega_d$, whereas for $m=0$, $R^m\kket{a,q+n_k}=P\kket{a,q+n_k}=0$ if $\kket{a,q+n_k}$ does not belong to the degenerate subspace, i.e. if $\tilde E_a\neq\tilde E_k+q\omega_d$.
As such, the terms with $m_j=0$ ($m_j>0$) contribute only for a resonant (non-resonant) $j$-th step, leading to the conditions in Eq.~\eqref{eq:constraint_m_0} and Eq.~\eqref{eq:constraint_m_neq0}.
}

The transformation~$W$ has a similar decomposition, with $m_0=0$ but $m_r$ not necessarily zero (corresponding to the components on the complementary subspace). Therefore, we obtain
\begin{align}\label{eq:W_expansion}
    W_r = \sum_{m_1+\dots+m_r=r}c^W_{m_r\dots,m_1}R^{m_r}\V\dots R^{m_1}\V P
\end{align}
with a different set of multiplicity coefficients $c^W_{m_r,\dots,m_1}$. 
At first and second order, we obtain
\begin{align}
    W_1 &= R\V P, \\ 
    W_2 &=  R\V R\V P - R^2\V P\V P - \frac{1}{2} P\V R^2\V P.
\end{align}
In particular, we have a single trivial coefficient at first order $c_1^W=1$,
and the coefficients at orders~2,~3 and~4 are listed in Table~\ref{table:DPTcoeff_W_order2-3} and~\ref{table:DPTcoeff_W_order4}.
\corr{
We expand the matrix elements of~$W_r$ in Eq.~\eqref{eq:W_expansion} similarly as outlined for the matrix elements of $\H_\eff^{(r)}$ above to obtain Eq.~\eqref{eq:W_mat_elmt}.
}

The above recurrence relations can be solved using a formal calculus programming language to obtain the coefficients $c_{m_{r-1},\dots,m_{1}}$ and $c^W_{m_r,\dots,m_1}$ at arbitrary order.
We developed such a numerical program written in Python and using the package SymPy. It is accessible on Zenodo~\cite{ZenodoCode} together with a list of the multiplicity coefficients up to order $r_H=12$ and $r_W=8$.
%
\begin{table}
\centering
\begin{tabular}{||c c | c||} 
 \hline
 $m_2$ & $m_1$ & $c_{m_2,m_1}$ \\
 \hline\hline
0 &	2 &	$-1/2$ \\
1 &	1 &	$1$ \\
2 &	0 &	$-1/2$ \\
 \hline
\end{tabular} \
\begin{tabular}{||c c c | c||} 
 \hline
 $m_3$ & $m_2$ & $m_1$ & $c_{m_3,m_2,m_1}$ \\
 \hline\hline
0 &	0 &	3 &	$1/2$ \\
0 &	1 &	2 &	$-1/2$ \\
0 &	2 &	1 &	$-1/2$ \\
1 &	0 &	2 &	$-1/2$ \\
1 &	1 &	1 &	$1$ \\
1 &	2 &	0 &	$-1/2$ \\
2 &	0 &	1 &	$-1/2$ \\
2 &	1 &	0 &	$-1/2$ \\
3 &	0 &	0 &	$1/2$ \\
 \hline
\end{tabular}
\caption{Non-zero multiplicity coefficients of $\H_\eff$ at order~3 and~4.}
\label{table:DPTcoeff_order3-4}
\end{table}
\begin{table}
\centering
\begin{tabular}{||c c c c | c||} 
 \hline
 $m_4$ & $m_3$ & $m_2$ & $m_1$ & $c_{m_4,m_3,m_2,m_1}$ \\
 \hline\hline
0 &	0 &	0 &	4 &	$-1/2$ \\
0 &	0 &	1 &	3 &	$1/2$ \\
0 &	0 &	2 &	2 &	$1/2$ \\
0 &	0 &	3 &	1 &	$1/2$ \\
0 &	1 &	0 &	3 &	$1/2$ \\
0 &	1 &	1 &	2 &	$-1/2$ \\
0 &	1 &	2 &	1 &	$-1/2$ \\
0 &	2 &	0 &	2 &	$3/8$ \\
0 &	2 &	1 &	1 &	$-1/2$ \\
1 &	0 &	0 &	3 &	$1/2$ \\
1 &	0 &	1 &	2 &	$-1/2$ \\
1 &	0 &	2 &	1 &	$-1/2$ \\
1 &	1 &	0 &	2 &	$-1/2$ \\
1 &	1 &	1 &	1 &	$1$ \\
1 &	1 &	2 &	0 &	$-1/2$ \\
1 &	2 &	0 &	1 &	$-1/2$ \\
1 &	2 &	1 &	0 &	$-1/2$ \\
1 &	3 &	0 &	0 &	$1/2$ \\
2 &	0 &	0 &	2 &	$1/4$\\
2 &	0 &	1 &	1 &	$-1/2$ \\
2 &	0 &	2 &	0 &	$3/8$ \\
2 &	1 &	0 &	1 &	$-1/2$ \\
2 &	1 &	1 &	0 &	$-1/2$ \\
2 &	2 &	0 &	0 &	$1/2$ \\
3 &	0 &	0 &	1 &	$1/2$ \\
3 &	0 &	1 &	0 &	$1/2$ \\
3 &	1 &	0 &	0 &	$1/2$ \\
4 &	0 &	0 &	0 &	$-1/2$ \\
 \hline
\end{tabular}
\caption{Non-zero multiplicity coefficients of~$\H_\eff$ at order~5.}
\label{table:DPTcoeff_order5}
\end{table}
\begin{table}
\centering
\begin{tabular}{||c c | c||} 
 \hline
 \rule{0pt}{10pt}
 $m_2$ & $m_1$ & $c_{m_2,m_1}^W$ \\
 \hline\hline
0 &	2 &	$-1/2$ \\
1 &	1 &	$1$ \\
2 &	0 &	$-1$ \\
 \hline
\end{tabular} \
\begin{tabular}{||c c c | c||} 
 \hline
 \rule{0pt}{10pt}
 $m_3$ & $m_2$ & $m_1$ & $c_{m_3,m_2,m_1}^W$ \\
 \hline\hline
0 & 0 & 3 & $1/2$ \\
0 & 1 & 2 & $-1/2$ \\
0 & 2 & 1 & $-1/2$ \\
0 & 3 & 0 & $1/2$ \\
1 & 0 & 2 & $-1/2$ \\
1 & 1 & 1 & $1$ \\
1 & 2 & 0 & $-1$ \\
2 & 0 & 1 & $-1$ \\
2 & 1 & 0 & $-1$ \\
3 & 0 & 0 & $1$ \\
 \hline
\end{tabular}
\caption{Non-zero multiplicity coefficients of $W$ at order 2 and 3.}
\label{table:DPTcoeff_W_order2-3}
\end{table}
\begin{table}
\centering
\begin{tabular}{||c c c c | c||} 
 \hline
 \rule{0pt}{10pt}
 $m_4$ & $m_3$ & $m_2$ & $m_1$ & $c_{m_4,m_3,m_2,m_1}^W$ \\
 \hline\hline
0 & 0 & 0 & 4 & $-1/2$ \\
0 & 0 & 1 & 3 & $1/2$ \\
0 & 0 & 2 & 2 & $1/2$ \\
0 & 0 & 3 & 1 & $1/2$ \\
0 & 0 & 4 & 0 & $-1/2$ \\
0 & 1 & 0 & 3 & $1/2$ \\
0 & 1 & 1 & 2 & $-1/2$ \\
0 & 1 & 2 & 1 & $-1/2$ \\
0 & 1 & 3 & 0 & $1/2$ \\
0 & 2 & 0 & 2 & $3/8$ \\
0 & 2 & 1 & 1 & $-1/2$ \\
0 & 2 & 2 & 0 & $1/2$ \\
0 & 3 & 0 & 1 & $1/2$ \\
0 & 3 & 1 & 0 & $1/2$ \\
0 & 4 & 0 & 0 & $-1/2$ \\
1 & 0 & 0 & 3 & $1/2$ \\
1 & 0 & 1 & 2 & $-1/2$ \\
1 & 0 & 2 & 1 & $-1/2$ \\
1 & 0 & 3 & 0 & $1/2$ \\
1 & 1 & 0 & 2 & $-1/2$ \\
1 & 1 & 1 & 1 & $1$ \\
1 & 1 & 2 & 0 & $-1$ \\
1 & 2 & 0 & 1 & $-1$ \\
1 & 2 & 1 & 0 & $-1$ \\
1 & 3 & 0 & 0 & $1$ \\
2 & 0 & 0 & 2 & $1/2$ \\
2 & 0 & 1 & 1 & $-1$ \\
2 & 0 & 2 & 0 & $1$ \\
2 & 1 & 0 & 1 & $-1$ \\
2 & 1 & 1 & 0 & $-1$ \\
2 & 2 & 0 & 0 & $1$ \\
3 & 0 & 0 & 1 & $1$ \\
3 & 0 & 1 & 0 & $1$ \\
3 & 1 & 0 & 0 & $1$ \\
4 & 0 & 0 & 0 & $-1$ \\
 \hline
\end{tabular}
\caption{Non-zero multiplicity coefficients of $W$ at order 4.}
\label{table:DPTcoeff_W_order4}
\end{table}

We remark that the transformation~$W$ can alternatively be derived from an anti-hermitian generator~$W=e^{S}$, where $S^\dagger=-S$~\cite{suzuki_degenerate_1983}, which is then equivalent to the Schrieffer-Wolff approach~\cite{schrieffer_relation_1966,bravyi_schriefferwolff_2011}.
The advantage of the current approach is that it allows us to obtain a systematic expansion of the effective Hamiltonian expressed only in terms of the resolvent operator~$R$, the degenerate projection~$P$, and the perturbation~$\V$.

\section{Sub-harmonic drive with Gaussian flat-top pulse}\label{app:pulse_shaping}

To address the observed leakages in large driving amplitudes and the fast oscillations, it is customary to introduce pulse shapes, i.e., vary the amplitude of the drive over time.
While the current constant amplitude results in a periodic Hamiltonian, allowing the use of Floquet theory, alternative pulse shapes have been shown to improve fidelity, such as ramping the amplitude up and down in the form of a Gaussian flatop~\cite{hertel_gate-tunable_2022}.
Intuitively, a smooth pulse will remove unwanted Fourier components that an instantaneous rectangular pulse would contribute, which leads to decreased fidelity. 
Further, a smooth pulse removes fine-tuning issues when for fast oscillations at large drive amplitude, where the resolution of gate times is limited by current AWG devices (typically allowing an accuracy of around 0.5ns).

More quantitatively speaking, we replace the periodic flux drive by an optimized flux pulse with a finite duration $T$, characterized by an envelope function $E(t)$ with $E(0)=E(T)=0$
\begin{align}
     H_d(t) &= - E_L A\, E(t)\, \cos(\omega_d t) \varphi
\end{align}
In particular, we choose the envelope function to be the Gaussian-flap-top envelopes with a piecewise definition as
\begin{align}
    E(t)=\begin{cases}
        \exp{-\frac{(t-t_0)^2}{2\sigma^2}} & 0\leq t < t_0 \\
        1 & t_0\leq x< T-t_0 \\
        \exp{-\frac{(t-T+t_0)^2}{2\sigma^2}} & T-t_0\leq t \leq T
    \end{cases}\;.
\end{align}
For the purpose of numerically verifying the following description with fluxonium qubits, we will adapt the standard deviation of the Gaussian-type envelope pulse to be $\sigma=4 \ \text{ns}$ and the risetime to be equal to the falltime as $t_0=18 \ \text{ns}$.

It is now possible - and customary in the quantum computing community - to use numerical optimization of the parameters $A,\omega_d$ to identify those for which the gate reaches maximum fidelity. This calibration process might be guided by the analytical findings of the previous section to identify suitable starting points. This is based on the expectation that small changes to the pulse shape will only slightly alter the resonance condition and Rabi rate. However, we emphasize that indeed both expression change and the highest fidelities can only be achieved if one finds the new modified expressions.

As an alternative, one may also try to get an analytical understanding of the changes to the resonance condition. A more general procedure will be a useful enhancement in situations where the brute-force optimization struggles to alleviate the numerical efforts. In the remainder of this section, we therefore establish a second perturbative framework, which determines improved resonance conditions and Rabi rates from non-periodic pulse shapes.

Conceptually, one may imagine a rise- and fall-phase of the pulse as creating an adiabatic evolution from the static qubit eigenstates $\ket{n}$ into the eigenstates of the Floquet Hamiltonian.
A possible way to envision this adiabatic evolution is by using the Floquet-Magnus expansions formalism, stated in~\cite{schirk_subharmonic_2025}, in which the  average dynamics is obtained by a time-dependent unitary transformation $e^{iK(t)}$, where $K(t)$ depends on $E(t)$.
Different from the bare qubit state, during the main time of the pulse, the Floquet eigenstate will experience the desired resonant dynamics of the effective Hamiltonian $\H_\mathrm{eff}$ and refocus to the qubit state at the rest frame.

Introducing a shaped envelope function on the drive will necessarily alter the resonance condition, obtained from $\delta_{1}^{[r_\mathrm{max}]}-\delta_{0}^{[r_\mathrm{max}]}=0$, which determines the resonant drive frequency $\omega_\mathrm{d}$. As the framework of Floquet theory is valid only in periodic systems, it is not possible to directly incorporate the changes to the resonance condition therein. However, we expect to obtain a first intuition on the effects in the limit of slowly changing envelope functions compared to the drive frequency. In this situation, the effective Hamiltonian of the reduced two-level system obtained by~\cref{eq:eff_fluxonium_hamiltonian} is only changing slowly, such that an additional Magnus-expansion \cite{magnus_exponential_1954} of the Hamiltonian becomes meaningful. In other words, we assume to work in the parameter regime:
\begin{align}
    2\pi/\omega_d \ll t_0 \ll \pi/\Omega_{10}
\end{align}
As higher orders of the Magnus expansion will only be non-trivial during rise- and fall-time, and should not accumulate to a half oscillation in order to guarantee convergence. In this way, the evolution decomposes into three parts: rise and fall with time-dependent effective Hamiltonian $\H_{\eff}(t)$ and a mostly time-independent part given by the usual $\H_{\eff}$ from degenerate Floquet perturbation theory:
\begin{equation}
\begin{aligned}
    U(T,0)&=U(T,T-t_0)U(T-t_0,t_0)U(t_0,0)\\
    &=e^{-S(t_0)}e^{-i(T-2t_0)\H_{\eff}} e^{S(t_0)}\\
    &=:\exp(-iT H_M)
\end{aligned}
\end{equation}
where the $S$ operator captures the ramping effects in the envelope functions, and is defined by 
\begin{align}
e^{S(t_0)}:=\mathcal{T}\exp{-i\int_{0}^{t_0}\dd{t} \H_{\eff}(t)} 
\end{align}
Within the approximation of adiabaticity, we then replace the driving amplitude in $\H_{\eff}$  by $A \rightarrow A E(t)$ for the rise and fall phase, leading to the time-dependent effective Hamiltonian 
\begin{align}
\mathcal{H}_{\eff}(t)&=-\frac{\delta_{1}(t)-\delta_{0}(t)}{2}\sigma_z +\Omega_{01}(t)\sigma_x \\
&=-\frac{\delta(t)}{2}\sigma_z +\Omega(t)\sigma_x
\label{eq:Ham_Pauli_represent}
\end{align}
Using the Magnus expansion, the $S(t_0)$ operator is explicitly $S(t_0) = \sum_{k=0}^\infty S_k$
with
\begin{align}
    S_0 &= \frac{1}{t_0} \int_0^{t_0}\dd{t} \H_{\eff}(t) \\ 
    S_1 &= \frac{1}{2t_0} \int_0^{t_0}\dd{t_1}\int_0^{t_1}\dd{t_2} \left[\H_{\eff}(t_1),\H_{\eff}(t_2)\right] \\
    &\cdots \nonumber
\end{align}
Finally, the Baker–Campbell–Hausdorff formula gives the time-independent generator $H_M$ as follows
\begin{align}
    H_{M}=    &\left(\H_\eff-[S(t_0),\H_\eff]+\frac{1}{2}[S(t_0),[S(t_0),\H_\eff]]+\cdots\right) \nonumber\\
    &\times\frac{T-2t_0}{T}
\end{align}
When plugging in \cref{eq:Ham_Pauli_represent}, we collect contributions in the number of nested commutators, which are generated by Pauli matrices as follows.
\begin{align}
    H_M =-\frac{\delta_{M}}{2}\sigma_z+\Omega_{M}^x\sigma_x+\Omega_{M}^y\sigma_y := \frac{T-2t_0}{T}\sum_k H_M^{(k)}
\end{align}
where
\begin{align}
    H_M^{(0)}&=\H_\eff=-\frac{\delta(t_0)}{2}\sigma_z +\Omega(t_0) \sigma_x \\
    H_M^{(1)}&=[S_0,\H_\eff]\nonumber\\
    &= -\frac{i\sigma_y}{t_0} \int_0^{t_0}  \delta(t)\Omega(t_0)-\delta(t_0)\Omega(t) \; \dd t\\
    H_M^{(2)}&=[S_1,\H_\eff]-\frac{1}{2}[S_0,[S_0,\H_\eff]] \label{eq:2ndHM}
\end{align}
The higher-order contributions from the BCH formula can be computed accordingly.

To evaluate this expression, we collect contributions from Floquet perturbation theory at different orders.
Explicitly, the time-dependent Stark shifts and the effective coupling in \cref{eq:Ham_Pauli_represent} are power series in driving amplitude $A$.
\begin{align*}
    \delta(t) & = \epsilon+\delta^{(2)}(\epsilon, t)A^2+\delta^{(3)}(\epsilon, t)A^3+\delta^{(4)}(\epsilon, t)A^4+\cdots\\
    \Omega(t) & = \Omega^{(3)}(\epsilon, t)A^3+\Omega^{(4)}(\epsilon, t)A^4+\Omega^{(5)}(\epsilon, t)A^5+\cdots
\end{align*}
For illustrative purposes, we have here extracted denoted the $k$th order coefficients in both expressions, i.e., $\delta^{(k)}$ and $\Omega^{(k)}$.

Thus, after integrating out rising- and falling-time $t_0$, one obtains a pair of equations by equaling the modified Stark shifts $\delta_M$ and setting its corresponding $\pi$-rotation 
\begin{align}
\label{eq:Envelop_conditions_stark_shifts}
\delta_{M}(\epsilon, A, T)&=0 \\
\Omega_{M}(\epsilon, A, T)&=\frac{\pi}{T-2t_0}
\label{eq:Envelop_conditions_rabi_rates}
\end{align}
where the modified effective coupling $\Omega^2_{M}=(\Omega^x_{M})^2+(\Omega^y_{M})^2$.
Here, for any given drive amplitude $A$, one can obtain the resonant drive frequency from the frequency detuning $\epsilon$ and characterize the Rabi rate from the gate time $T$ of the $\pi$-pulse.

\begin{figure}
    \centering
    \includegraphics[width=.99\linewidth]{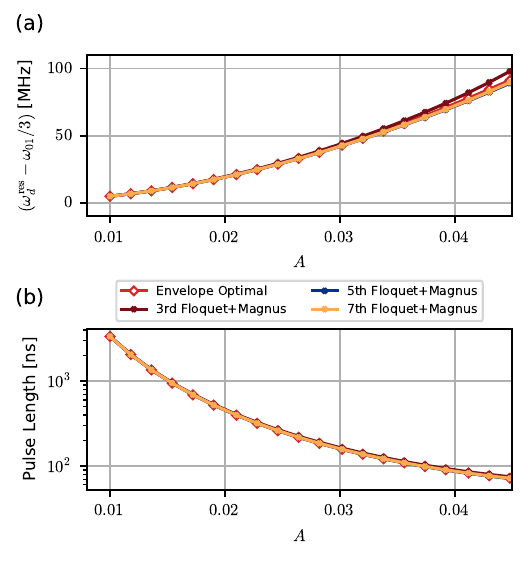}
    \caption{\textbf{Comparing predictivity at different orders with pulse shaping.} The frequency shift {\bf (a)} and duration of $\pi$ pulse  {\bf (b)} for driving Rabi oscillations with sub-harmonic processes are numerically determined (red dots) for various drive amplitudes. The predictions combining degenerate Floquet-perturbation theory at finite orders (solid colours) and Magnus expansions at the 3rd order show a convergence towards the numerical values, especially fast at low drive amplitudes.
    Note that the theoretical predictions estimate gate time $T$ by solving~\cref{eq:Envelop_conditions_stark_shifts,eq:Envelop_conditions_rabi_rates}.
    }
    \label{fig:pulse_shaping}
\end{figure}

After considering in this way the effects of the envelope function, we terminate the BCH expansion as well as the Magnus expansion at, e.g., the 3rd order. We overlap the predictions of frequency shifts $\epsilon$ as well as $\pi$-rotation time from Floquet theory at finite orders (solid colours) with numerical optimization (red dots) in~\cref{fig:pulse_shaping} (a) and (b). As expected from perturbative computations, we observe that for large driving amplitudes, the finite order predictions deviate from numerically optimized gates. However, with increasing perturbation order, the predictions improve towards the numerical results.

To quantitatively characterize the effects of the enveloped pulse, we compare the errors among different models in~\cref{fig:pulse_shaping_fidelity}. At small amplitudes, errors decrease when going to Floquet perturbation theory at higher orders, as the 5th and 7th orders reached $10^{-5}$, nearly converged, and are better than predictions from the 3rd order.
A key observation is that when going to larger driving amplitudes $A/2\pi\approx0.05$, theoretical predictions, even at higher orders, lead to large errors around $10\%$. This is because it reaches the boundary of the converging regime in the Magnus expansion. The Magnus expansion converges whenever~\cite {magnus_exponential_1954}
\begin{equation}
    \int_0^{t_0} ||\H_\eff(t)||_2 \dd t = \int_0^{t_0} \Omega(t) \dd t < \log 2
\end{equation}
where we assume we are close to resonant conditions so that contributions from Stark shifts can be neglected. 

As anticipated, the use of a Gaussian flat-top pulse significantly improves fidelity compared to a constant pulse. 
This is due to its ability to precisely tune both the fast oscillations and leakages to higher states. 
Our results indicate that the predictions from $r_H=7$ perturbation theory (solid yellow) achieves infidelities below \(10^{-5}\) with small amplitudes, \(A/2\pi < 0.025\). 
This level of infidelity is lower than what is observed with the best-case periodic drive (grey dots), which maintains the same optimal infidelity as depicted in~\cref{fig:comparison} (c) as a comparison.
Furthermore, we strengthen that one could achieve infidelities to be $10^{-7}$ with enveloped functions (red dots), showing huge potential in optimal control of sub-harmonic driving.

\section{Comparison to other approximations within the transmon regime}
\label{app:transmon}

Previously, sub-harmonic driving has been applied to the transmon qubit~\cite{xia_fast_2025} at $n_1=3$rd order. The model of~\cite{xia_fast_2025} used a frame change to extract the resonant frequency shifts and the Rabit rates utilizing a RWA~\cite{xia_fast_2025}. To compare the accuracy of this model to the degenerate perturbation theory put forward in this paper, we shortly repeat the main steps, referring to \cite{xia_fast_2025} for all details. A direct coupling between the external source and a single transmon qubit gives the driven transmon Hamiltonian
\begin{align}
    \begin{aligned}
    H_\mathrm{transmon}=(\omega_\mathrm{q}-\alpha)a^\dagger  a&+\frac{\alpha}{12}(a^\dagger + a)^4 \\
    &+A\cos(\omega_dt)\left(a+a^\dagger\right),
\end{aligned}
\label{eq:driven-qubits model}
\end{align}
where typical experimental parameters are $\omega_{\rm q}/2\pi\approx 3-7$ GHz and an anharmonicity of $\alpha/\pi\in[-0.2\, {\rm GHz},-0.4 \,{\rm GHz}]$. In Eq.~\eqref{eq:driven-qubits model}, $a$ denotes the ladder operator of a usual harmonic oscillator with $[a,a^\dagger]=1$. To see the three-photon transition, \cite{xia_fast_2025} applied a displacement operator
\begin{align}
    D_z(t)=\exp{za^\dagger -z^* a}, 
\end{align}
where
\begin{align}
    z=-\frac{A}{2(\omega_d-[\omega_\mathrm{q}-\alpha)]}e^{-i\omega_dt}+\frac{A}{2[\omega_d+(\omega_\mathrm{q}-\alpha)]}e^{i\omega_dt}\;.
\end{align}
Upon expanding the fourth power and applying a RWA, the driven Hamiltonian after the frame change reduces to 
\begin{align}
    H_{FC}=(\epsilon+2\alpha |\eta|^2)a^\dagger  a +\frac{\alpha}{2} a^\dagger a^\dagger a a+\frac{\alpha\eta^3}{3}(a^\dagger + a)
\end{align}
with $\eta=\frac{(\omega_\mathrm{q}-\alpha)A}{\omega_d^2-(\omega_\mathrm{q}-\alpha)^2}$ and $ \epsilon = \omega_\mathrm{q}-3\omega_d$.
Thus, the relevant detunings and the effective coupling rates that follow from this analysis are given by
\begin{align}
    &\delta_{1,FC}-\delta_{0,FC}=\epsilon+2\alpha|\eta|^2=\epsilon +\frac{2\alpha(\omega_\mathrm{q}-\alpha)^2A^2}{(\omega_d^2-(\omega_\mathrm{q}-\alpha)^2)^2}, \label{eq:transmon_shifts}\\&\Omega_{10,FC}=\frac{\alpha \eta^3}{3}=\frac{\alpha(\omega_\mathrm{q}-\alpha)^3A^3}{3(\omega_d^2-(\omega_\mathrm{q}-\alpha)^2)^3}.\label{eq:transmon_Rabi}
\end{align}

\begin{figure}
    \centering
    \includegraphics{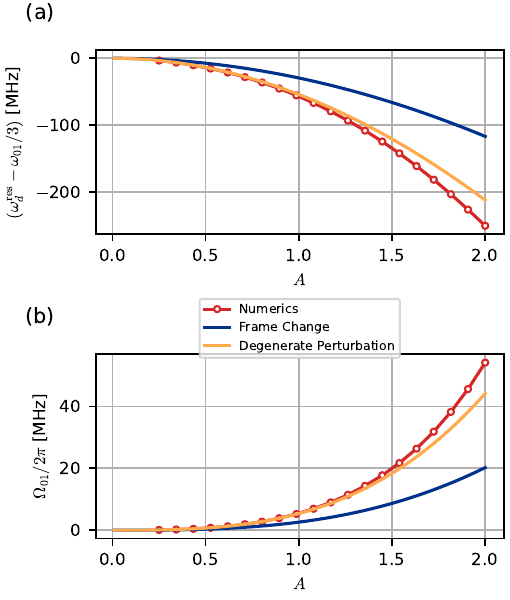}
    \caption{\textbf{Comparing predictivity of different approximations for sub-harmonically driven transmon.}
    The frequency shift (red)  and Rabi rates (blue) for a sub-harmonically driven transmon qubit are numerically determined for various drive amplitudes. The Floquet perturbation theory at the 3rd order (solid lines) predicts the numerical simulation at all values of driving amplitude better than the ``rotating-wave'' model (dashed lines) by \cite{xia_fast_2025}. 
    }
    \label{fig:comparison_transmon}
\end{figure}

Conversely, we can also extract the effective Hamiltonian for the transmon directly as a special case of the previously derived fluxonium Hamiltonian. As already outlined in the previous section, there is a marginal difference between Floquet-Magnus and degenerate Floquet perturbation theory. Hence, we will adapt equation~\cref{eq:driven-qubits model} by diagonalizing the transmon Hamiltonian, leaving a slight nonlinear drive term, i.e., $\beta_0\approx1,\beta_1\approx0$
\begin{equation}
    \begin{aligned}
    H=\omega'_\mathrm{q} a^\dagger a &+ \frac{\alpha'}{2} a^\dagger a^\dagger a a + A \beta_0 \cos(\omega_dt) (a^\dagger + a) \\
    &+A \beta_1 \cos(\omega_dt) (a^\dagger a^\dagger a + a^\dagger a a).
\end{aligned}
\end{equation}
Using \cref{eq:effective_energy_order2_monochrom} from Floquet perturbation theory, one can derive the lowest 2nd-order resonant frequency shifts.
To show the difference in the previous transmon model, we collect $\omega_\mathrm{q}'-\alpha'$ term in the denominator and apply the Taylor expansions around small $\alpha$ with termination in linear terms
\begin{align}
    \delta_{1,DP}^{[3]}-\delta_{0,DP}^{[3]}
    = \epsilon +\frac{2\alpha[(\omega_\mathrm{q}'-\alpha')^2+\omega_d^2]A^2}{[\omega_d^2-(\omega_\mathrm{q}'-\alpha')^2]^2} +\mathcal{O}(\alpha'^2).
\end{align}
Compared to the results in~\cref{eq:transmon_shifts}, we observe extra terms proportional to $\omega_d^2$ in the nominator. 
As for the coupling rate, one could collect terms and apply the Taylor expansions as well, with
\begin{align}
    \Omega_{10,DP}^{[3]}\approx -\frac{8\alpha  (\omega_\mathrm{q}'-\alpha')^3 A^3}{27[\omega_d^2-(\omega_\mathrm{q}'-\alpha')^2]^3 } +\mathcal{O}(\alpha'^2),\label{eq:transmon_FM_Omega}
\end{align}
where we find the same scaling as in~\cref{eq:transmon_Rabi}, but a drastically different prefactor.

The physical origin of these deviations can be traced back to an accumulation of various effects that the RWA neglects: higher order correction terms originating from the size of the anharmonicity, the speed of the drive frequency, and the effects of higher levels beyond a two-level approximation. 

To conclude, we compare whether the determined difference in terms of predictions between the two models in the transmon qubit is relevant for experimentally accessible regimes. For this purpose, we take the parameters of the particular transmon defined in~\cite{xia_fast_2025}, $\omega_\mathrm{q}/2\pi=3.96 \; \text{GHz}, \; \alpha/2\pi=-208 \; \text{MHz}$ and extract the Stark shift with the Rabi Rates in~\cref{fig:comparison_transmon} (a). It is clearly visible that the experimental values for both resonance conditions and Rabi rate are largely off, even at low values of drive amplitude. In numerical simulations, we observed that transition processes induced by the 3-photon sub-harmonic driving exist up to $A/2\pi =5 \; \text{GHz}$. However, amplitudes larger than $ 2 \; \text {GHz}$ show lower than $75\%$ fidelity in the single qubit gate, as is shown in ~\cref{fig:comparison_transmon} (b), due to the large leakage into higher lying states caused by  the small anharmonicity of the transmon.

Speaking in practical terms, for one-qubit gates, the model predictions will only serve as a starting point for a closed-loop optimizer on the experimental devices, hence leading to similarly good results for both models. However, it is expected that upon scaling to larger systems, the enhanced predictivity and generality of the perturbation theory developed in this manuscript will become relevant.

\end{document}